\documentclass[12pt]{article}
\usepackage{epsf,epsfig,cite}
\newcount\timecount
\newcount\hours \newcount\minutes  \newcount\temp \newcount\pmhours

\hours = \time
\divide\hours by 60
\temp = \hours
\multiply\temp by 60
\minutes = \time
\advance\minutes by -\temp
\def\hour{\the\hours}
\def\minute{\ifnum\minutes<10 0\the\minutes
            \else\the\minutes\fi}
\def\clock{
\ifnum\hours=0 12:\minute\ AM
\else\ifnum\hours<12 \hour:\minute\ AM
      \else\ifnum\hours=12 12:\minute\ PM
            \else\ifnum\hours>12
                 \pmhours=\hours
                 \advance\pmhours by -12
                 \the\pmhours:\minute\ PM
                 \fi
            \fi
      \fi
\fi
}

\def\monthname{\relax\ifcase\month 0/\or January\or February\or
   March\or April\or May\or June\or July\or August\or September\or
   October\or November\or December\else\number\month/\fi}

\def\bold#1{\setbox0=\hbox{$#1$}%
     \kern-.025em\copy0\kern-\wd0
     \kern.05em\copy0\kern-\wd0
     \kern-.025em\raise.0433em\box0 }

\def\gappeq{\mathrel{\rlap {\raise.5ex\hbox{$>$}}
{\lower.5ex\hbox{$\sim$}}}}

\def\lappeq{\mathrel{\rlap{\raise.5ex\hbox{$<$}}
{\lower.5ex\hbox{$\sim$}}}}

\def\la{\mathrel{\raise.3ex\hbox{$<$\kern-.75em\lower1ex\hbox{$\sim$}}}}
\def\beq{\begin{equation}}
\def\eeq{\end{equation}}

\def\ohsq{\Omega_{\chi} h^2}

\def\m12{m_{1\!/2}}

\input paperdef

\begin{document}
\begin{titlepage}
\pagestyle{empty}
\baselineskip=21pt
\rightline{hep-ph/0211206}
\rightline{CERN--TH/2002-289, DCPT/02/122}
\rightline{IPPP/02/61, LMU 09/02}
\rightline{UMN--TH--2114/02, TPI--MINN--02/43}
\vskip 0.05in
\begin{center}
{\large{\bf Precision Analysis of the Lightest MSSM Higgs Boson\\
at Future Colliders}}
\end{center}
\begin{center}
\vskip 0.05in
{{\bf John Ellis}$^1$, 
{\bf Sven Heinemeyer}$^2$,
{\bf Keith A.~Olive}$^{3}$
and {\bf Georg Weiglein}$^{4}$}\\
\vskip 0.05in
{\it
$^1${TH Division, CERN, Geneva, Switzerland}\\
$^2${Institut f\"ur theoretische Elementarteilchenphysik,
LMU M\"unchen, Theresienstr.\ 37, D-80333 M\"unchen, Germany}\\
$^3${Theoretical Physics Institute, School of Physics and Astronomy,\\
University of Minnesota, Minneapolis, MN~55455, USA}\\
$^4${Institute for Particle Physics Phenomenology, University of Durham,\\
Durham DH1~3LE, UK}\\
}
\vskip 0.1in
{\bf Abstract}
\end{center}
\baselineskip=18pt \noindent


We investigate the sensitivity of observables measurable in 
$e^+ e^-$, $\ga \ga$ and $\mu^+ \mu^-$ collisions for distinguishing
the properties of the light neutral
$\cp$-even Higgs boson in the minimal supersymmetric extension of the
Standard Model (MSSM) from those of a Standard Model (SM) Higgs boson 
with the same
mass. We explore first the available parameter space in the constrained
MSSM (CMSSM), with universal soft supersymmetry-breaking parameters,
incorporating the most recent direct limits on sparticle and Higgs masses,
the indirect constraints from $b \to s \ga$ and $g_\mu - 2$, and the
cosmological relic density $\ohsq$. We calculate the products of the
expected CMSSM Higgs production cross sections and decay branching ratios
$\sigma \times \cB$ normalized by the corresponding values expected for 
those of a SM Higgs boson of the same mass. The results are compared with 
the
precisions expected at each collider, and allow for a direct comparison of
the different channels.  
The measurements in the Higgs sector are found to provide important
consistency tests of the CMSSM.
We then generalize our analysis to the case of a non-universal Higgs
model (NUHM), where the values of $\MA$ and $\mu$ are independent parameters.
We explore in particular the sensitivity to $\MA$, finding that measurements 
at $e^+ e^-$, $\ga \ga$ and $\mu^+ \mu^-$
colliders could yield indirect constraints on its value, up to 
$\MA \sim 1200 \gev$. 
We discuss the potential of these measurements for distinguishing
between the CMSSM and the NUHM, probing in this way the
assumption of universality.

\end{titlepage}


\section{Introduction}

In a previous paper~\cite{ehow}, we discussed the observability at the
Tevatron and the LHC of the lightest neutral Higgs boson in the
constrained MSSM (CMSSM)~\footnote{For a comparison of the Higgs-sector
properties of the CMSSM with 
gauge- and anomaly-mediated (GMSB and AMSB) scenarios, see~\cite{asbs}.}, 
in which the soft supersymmetry-breaking
parameters are assumed to be universal at some high GUT input
scale~\footnote{An economical way to ensure this universality is by
gravity-mediated supersymmetry breaking in a minimal supergravity (mSUGRA)  
scenario, but there are other ways to validate the CMSSM assumptions,
including no-scale supergravity scenarios.}. Our conclusions were in
general quite encouraging, in the sense that the products of hadronic
production cross sections and branching ratios $\si \times \cB$ in the 
CMSSM differ little
from those in the Standard Model (SM), so that a CMSSM Higgs boson should be
essentially as observable at the Tevatron or the LHC as would be a
SM Higgs boson with the same mass. 

On the other hand, the expected precision in measuring the Higgs boson
properties was found to be too small to 
establish deviations in the properties of the lightest CMSSM
Higgs boson from a SM Higgs boson with the same mass. As an example, the
decay $h \to \ga \ga$, which is the prime discovery channel for a CMSSM or
SM Higgs boson weighing $\sim 120 \gev$, has been analyzed
in~\cite{ehow}.  The statistical error is expected to be $\sim 1$~\% and
the parton-parton luminosity error about 5~\%. If the theoretical error in
the calculation of the parton-parton cross section could be neglected,
there could be a 2-$\sigma$ difference between the strengths of the CMSSM
and SM signals.  However, this may well be masked by the theoretical error
in the cross-section calculation, which is currently $\gsim
20$~\%~\cite{robi}, so the LHC may not be able to discriminate between
CMSSM and SM Higgs bosons. Thus the onus may fall on a subsequent lepton
collider to discriminate between them.

If supersymmetry as a low-energy theory is realized in nature, it is
likely that supersymmetric particles will be detected at the LHC and
future lepton colliders~\cite{Battaglia:2001zp}. While the observation of
supersymmetric particles would of course rule out the SM, it would
nevertheless be crucial to establish also that the Higgs sector has the
properties predicted within the MSSM. This holds in particular if only one
light Higgs boson which resembles the SM one is observed at the LHC. A
direct measurement of the heavy Higgs boson states of the MSSM might be
difficult or impossible, depending on $\tb$. For example, at the LHC there
is a wedge in the $(\MA, \tb)$ plane where the heavy Higgs bosons cannot
be detected, and direct observability at a lepton collider is limited by
the available centre-of-mass energy: see, e.g., \cite{eennH}. Even in the
case where additional Higgs bosons are observed, the precise measurements
of the properties of the lightest $\cp$-even Higgs boson will provide
important consistency tests of the model.

In this paper, we consider and compare the prospects of $e^+ e^-$ (LC), 
$\ga\ga$ (\gaC) and $\mu^+ \mu^-$ (\muC) colliders for establishing the
supersymmetric nature of the lightest $\cp$-even Higgs boson, as compared
to the properties a SM Higgs boson of the same mass would have.
We consider the principal Higgs observation channels at each collider,
see Table~\ref{tab:channels}, including their respective anticipated
accuracies~\cite{teslatdr,talkbrient,gaC1,gaC2,muC1}.
%
\begin{table}[htb]
\begin{center}
\renewcommand{\arraystretch}{1.5}
\begin{tabular}{|c|l|l|c|} \hline\hline
collider & production mode & decay mode & precision \\ \hline \hline
LC   & $e^+e^- \to Z^* \to Zh$ & $\hbb$     &  1.5\% \\ \hline
LC   &                         & $\htautau$ &  4.5\% \\ \hline
LC   &                         & $\hcc$     &  6\%   \\ \hline
LC   &                         & $\hgg$     &  4\%   \\ \hline
LC   &                         & $\hWW$     &  3\%   \\ \hline \hline
\gaC & $\ga \ga \to h$         & $\hbb$     &  2\%   \\ \hline
\gaC &                         & $\hWW$     &  5\%   \\ \hline 
\gaC &                         & $\hgaga$   &  11\%  \\ \hline \hline
\muC & $\mu^+\mu^- \to h$      & $\hbb$     &  3\%   \\ \hline \hline
\end{tabular}
\renewcommand{\arraystretch}{1.0}
\end{center}
\caption{\it Expected precisions in the measurements of Higgs observables 
at the LC, the \gaC\ and the \muC\ for the light $\cp$-even Higgs boson
of the MSSM. The production mode of the LC refers to running at
energies of $\sqrt{s} = 350$--$500 \gev$.}
\label{tab:channels}
\end{table}
%
In each case, we calculate the strength expected for a CMSSM Higgs signal
normalized relative to the SM signal
\BE
\frac{\Bigl[\si \times \cB \Bigl]_{\rm CMSSM}}
     {\Bigl[\si \times \cB \Bigl]_{\rm SM}},
\EE
as evaluated in~\cite{hff,hdecay}. We display our results in planes of 
the universal soft supersymmetry-breaking gaugino mass $m_{1/2}$ and 
scalar mass $m_0$,
for different representative values of $\tb$, the trilinear soft 
supersymmetry-breaking parameter $A_0$ and the sign of
the supersymmetric Higgs parameter $\mu$. In each case, we restrict our 
attention
to the regions of parameter space still permitted by the direct search
limits on sparticle~\cite{SUSYsearches} and Higgs
masses~\cite{LEPHiggs}, the indirect constraints from 
$b \to s \ga$~\cite{bsgexp,bsgtheory}, $g_\mu - 2$~\cite{g-2}, and the 
cosmological relic density $\ohsq$~\cite{EHNOS}, which we require to lie 
between 0.1 and 0.3~\cite{cdmexp}.
In this way, we identify the regions of the CMSSM parameter space in
which a certain channel may distinguish between CMSSM and SM Higgs bosons 
with the same mass. 

The mass of the $\cp$-odd Higgs boson, $\MA$, plays a key role in the
investigation of any MSSM scenario. Since, as discussed above, the direct
measurement of $\MA$ might be very difficult, it will be very important to
obtain indirect information about this parameter that can be confronted
with the predictions of the CMSSM or other supersymmetry-breaking
scenarios.  Specifically, we analyze a scenario in which the assumption
made in the CMSSM of universality between the soft supersymmetry-breaking
masses of the Higgs multiplets and those of the squarks and sleptons is
relaxed, a framework we term the non-universal Higgs model
(NUHM)~\cite{CMSSMuni}. Such non-universality releases $\mu$ and $\MA$
from the values that are fixed for them in the CMSSM, while otherwise the
spectrum of the supersymmetric particles resembles the one in the CMSSM.  
Since in the decoupling limit, $\MA \gg \MZ$, the Higgs sector of the MSSM
becomes SM-like, deviations in the production and decay of the lightest
$\cp$-even Higgs boson of the MSSM from the SM values can be translated
into an upper bound on $\MA$. Therefore, we seek in this paper to identify
the regions of the MSSM parameter space in which an indirect limit on
$\MA$ can be obtained from $h$~measurements alone, even if the $A$~boson
cannot be observed directly.

The sensitivity of the Higgs sector observables to variations in $\MA$
allows one to test the universality assumption of the CMSSM. 
Precise determinations of $\si \times \cB$ can in this way be used to
distinguish between the CMSSM and the NUHM. We investigate the potential
of the different colliders for setting limits on the deviations of $\mu$
and $\MA$ from their CMSSM values.

The rest of the paper is organized as follows. In
\refse{sec:constraints}, we summarize the CMSSM parameter space and its
phenomenological constraints. In \refse{sec:CMSSMvsSM}, we discuss
in which channel the lightest CMSSM Higgs boson can best be distinguished
from a SM Higgs boson with the same mass, at various accelerators. The 
indirect reach in $\MA$ and possible tests of the CMSSM universality
assumption are discussed in \refse{sec:MAreach}
Our conclusions are presented in \refse{sec:conclusions}.


\section{Phenomenological Constraints}
\label{sec:constraints}

Before describing our results in detail, we first review our treatment
of the experimental and cosmological constraints on the CMSSM parameter
space. 

Our treatment of the direct LEP constraints on sparticle masses is
described in~\cite{ehow}, so we do not describe it further here. The LEP
Higgs constraint within the SM is now $\mH > 114.4 \gev$, while the data 
show a 1.7-$\si$ excess over the background expectation compatible with
a Higgs signal with mass $\sim 116 \gev$~\cite{LEPHiggs}. As pointed
out previously, the $ZZh$ coupling in the CMSSM is very close to
that of the SM Higgs for almost all possible parameter values 
(see~\cite{asbs}, however), so the SM Higgs boson mass limit can be
carried over to the CMSSM for most of the 
parameter space. In this paper, we allow only CMSSM parameter choices that
are consistent with $\mh > 113 \gev$ as calculated for $m_t = 175 \gev$
using the {\tt FeynHiggs} code~\cite{fh,mhiggsFD2l} in its latest
implementation~\cite{mhiggsAEC}, which includes various recent
results~\cite{feynhiggs1.2,mhiggsalt2,mhiggsalsalb}.
We have chosen a somewhat weaker limit than the actual SM exclusion bound, 
owing to remaining theoretical uncertainties from unknown higher-order
corrections~\cite{mhiggsAEC}. The measured value of $\mh$, which
experimentally will be known with high precision in this scenario, 
will provide very valuable consistency tests of the model, provided that 
the theoretical uncertainties in the $\mh$ prediction can be reduced
below the level of about $1 \gev$. Thus, besides the bound $\mh > 113 \gev$,
for reference we also include in our plots the contours 
$\mh = 115, 117, 120, 125 \gev$.
In view of the experimental bounds on $\mh$, 
we do not consider values of $\tb$ below 10, since in the CMSSM the
low-$\tb$ region is severely constrained by the experimental bound
on the Higgs-boson mass.

In our treatment of $b \to s \ga$, we follow~\cite{ehow,EFGOSi} in our
implementation of NLO QCD corrections at large
$\tb$~\cite{bsgtheory}. 
We assume the 95\% confidence-level range 
$2.33 \times 10^{-4} < \cB(b \to s \ga) < 4.15 \times 10^{-4}$~\cite{bsgexp}, 
and we accept all CMSSM parameter sets that give 
predictions in this range, allowing for the scale and model dependences of
the QCD calculations.

The situation with regard to $a_\mu \equiv (g_\mu - 2)/2$ has changed
significantly since~\cite{ehow}. Concerning the theory evaluation, the
light-by-light contribution has been corrected~\cite{g-2lbl}.
Concerning the experimental precision, a new result has recently
been released by the E821 collaboration, including the year 2000
data~\cite{g-2}, which lead to a reduction in the experimental error
by roughly a factor of~2.
The magnitude of the deviation from the SM result is now at
$\delta a_\mu = (33.9 \pm 11.2) \times 10^{-10}$~\cite{Davier} 
using $e^+ e^-$ data for the hadronic vacuum polarization contribution
in the SM prediction, and by 
$\delta a_\mu = (16.7 \pm 10.7) \times 10^{-10}$~\cite{Davier} based on
$\tau$ decay data. Other recent analyses of the $e^+ e^-$ data yield
similar results~\cite{others}. We take the 2-$\si$ range to be
$11.5 \times 10^{-10} < \delta a_\mu < 56.3\times 10^{-10}$. 
This means that $\mu < 0$
is no longer allowed. In the following plots, we display as solid,
thick, red diagonal lines the $\pm 2$-$\si$ contours in the 
$(m_{1/2}, m_0)$ plane. However, we also show as thin red lines the
results of the more conservative theoretical estimate (based on the $\tau$ data)
$-4.7 \times 10^{-10} < \delta a_\mu < 38.1 \times 10^{-10}$, which allows some 
regions of parameter space with $\mu < 0$. 

As in~\cite{ehow}, we assume $R$-parity conservation, so that the
lightest supersymmetric particle (LSP), presumed to be the lightest
neutralino $\chi$, is stable and may have an interesting cosmological
relic density $\ohsq$. We accept CMSSM parameter sets that have $0.1 \le
\ohsq \le 0.3$ as calculated using the code documented
in~\cite{EFGOSi,Battaglia:2001zp}. 
Lower values of $\ohsq$ would be allowed if not all the cosmological dark
matter is composed of neutralinos. However, larger values of $\ohsq$ are
excluded by cosmology, and even values as large as 0.3 are disfavoured by
the most recent global fits to cosmological data~\cite{cdmexp,Melchiorri}.


\section{Sensitivity of Higgs-sector observables to deviations of
the CMSSM from the SM}
\label{sec:CMSSMvsSM}

In this Section we complete the survey of Higgs production and decay
channels, which we started for hadron colliders in~\cite{ehow}.  Here we
present analogous results for the LC, the \gaC\ and the \muC\ for the
channels given in Table~\ref{tab:channels}.  We present our results in
$(m_{1/2}, m_0)$ planes for $\tb = 10$, $\mu > 0$ and $A_0 = 0$, for $\tb
= 50$, $\mu > 0$ and $A_0 = 0$ or $-2 m_{1/2}$, and for $\tb = 35$, $\mu <
0$ and $A_0 = m_{1/2}$.  The irregularities in the cosmological region in
panel (a) etc., and the separations between the dots in panel (b) etc.\ are
due to the finite grid size used in our sampling of parameter space.  The
figures are provided with $\MA$ contours, showing the sensitivity of each
channel to this fundamental parameter of the Higgs boson sector.  The
resultant indirect constraints on $\MA$ are discussed in 
Section~\ref{sec:MAreach}.


\subsection{Observables at an $e^- e^+$ Linear Collider}
\begin{figure}[!ht]

\begin{center}
\sibrfrac{\eeZh}{\hbb}
\end{center}

\vspace{-1em}

\begin{minipage}{8in}
\hspace{4cm} (a) \hspace{7.2cm} (b) 
\end{minipage}

\vspace{.2em}

\begin{minipage}{8in}
\epsfig{file=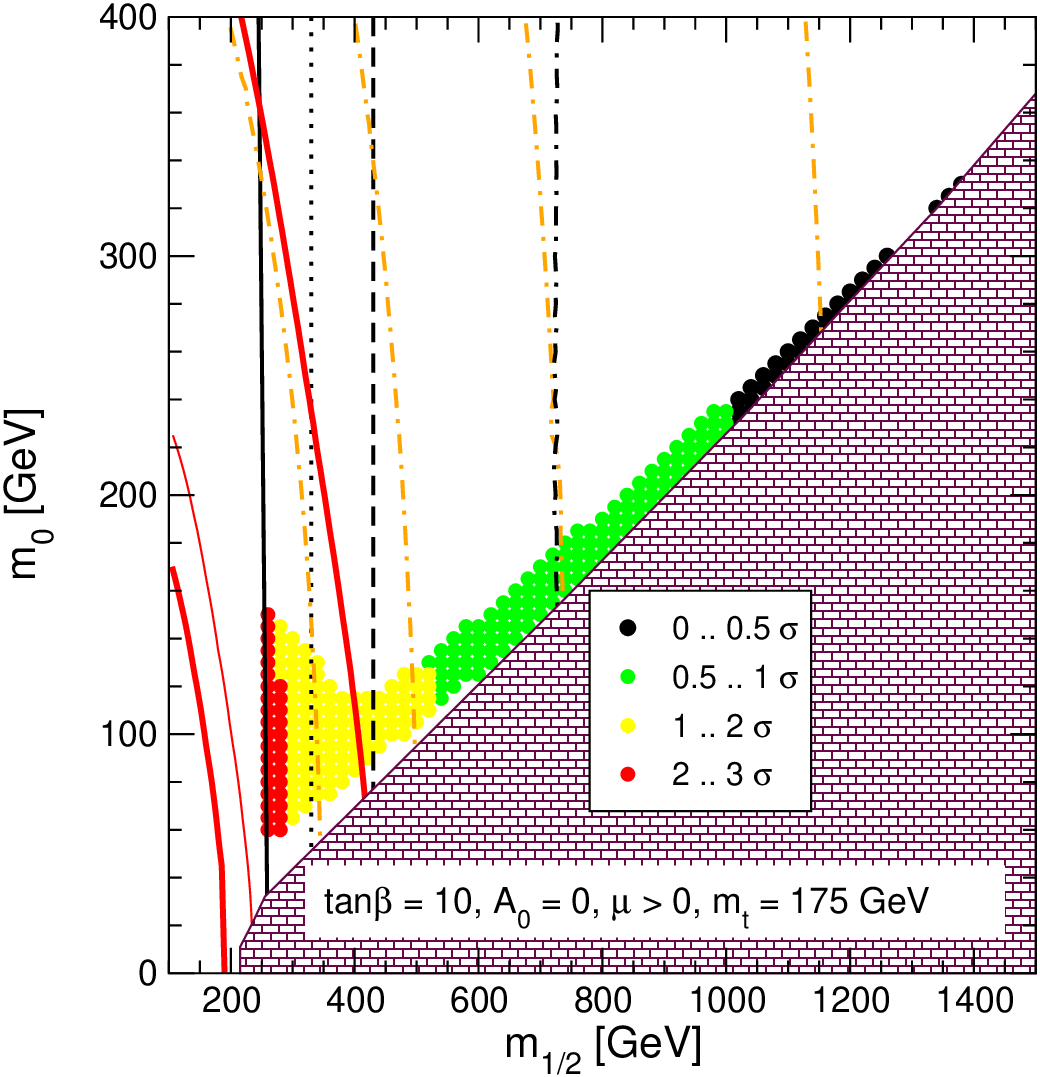,height=3.2in}
\epsfig{file=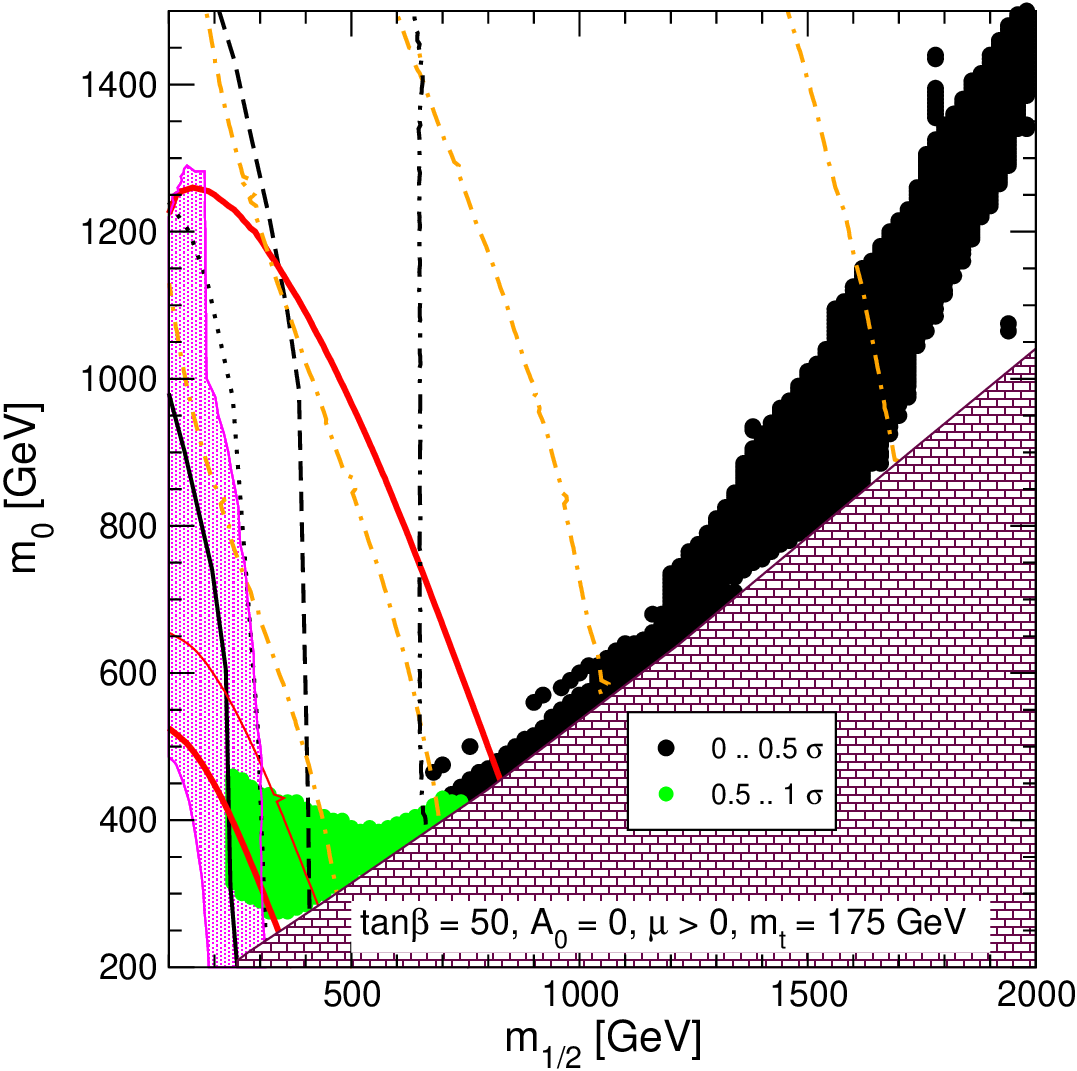,height=3.2in} \hfill
\end{minipage}

\vspace{.5em}

\begin{minipage}{8in}
\hspace{4cm} (c) \hspace{7.2cm} (d) 
\end{minipage}

\vspace{.2em}

\begin{minipage}{8in}
\epsfig{file=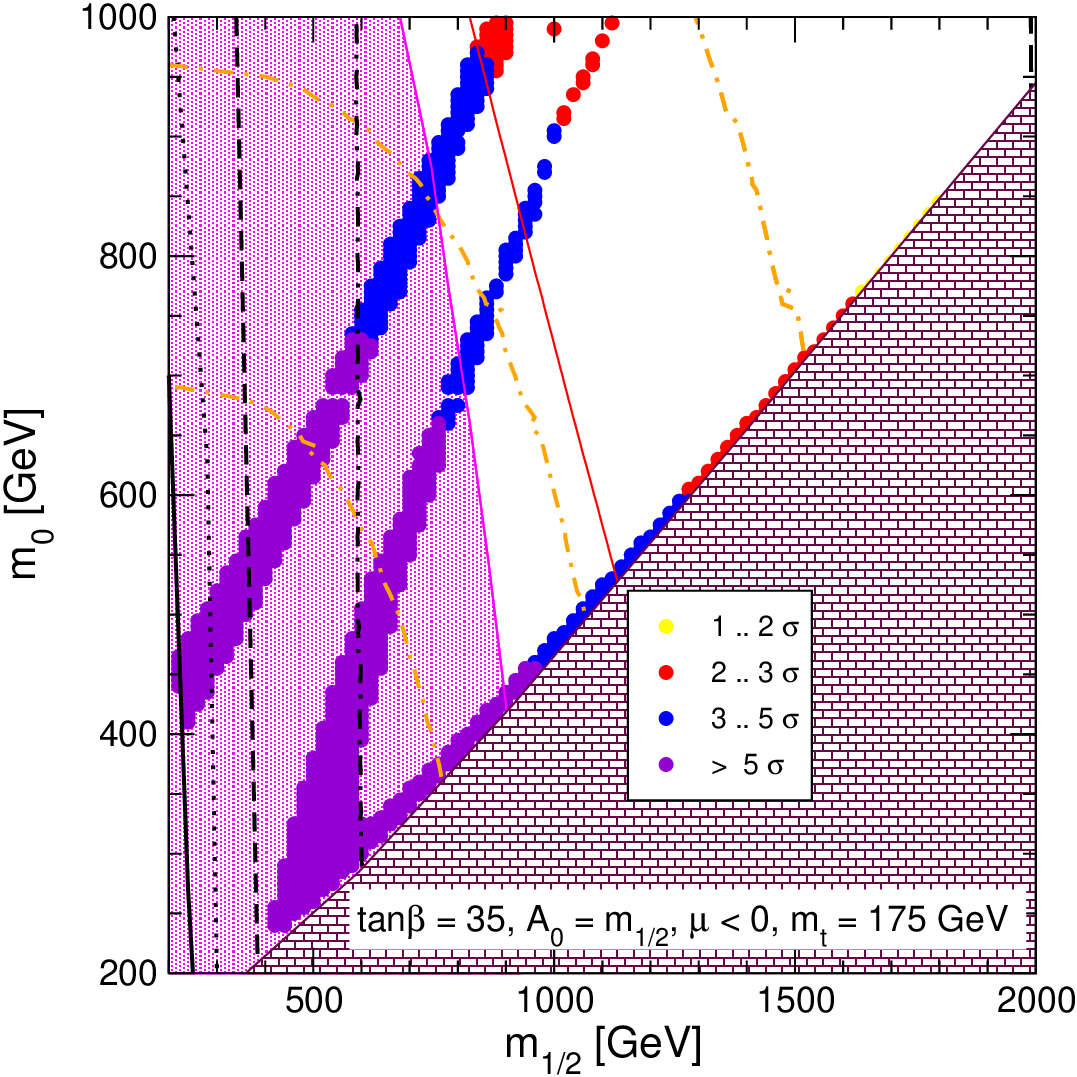,height=3.2in}
\epsfig{file=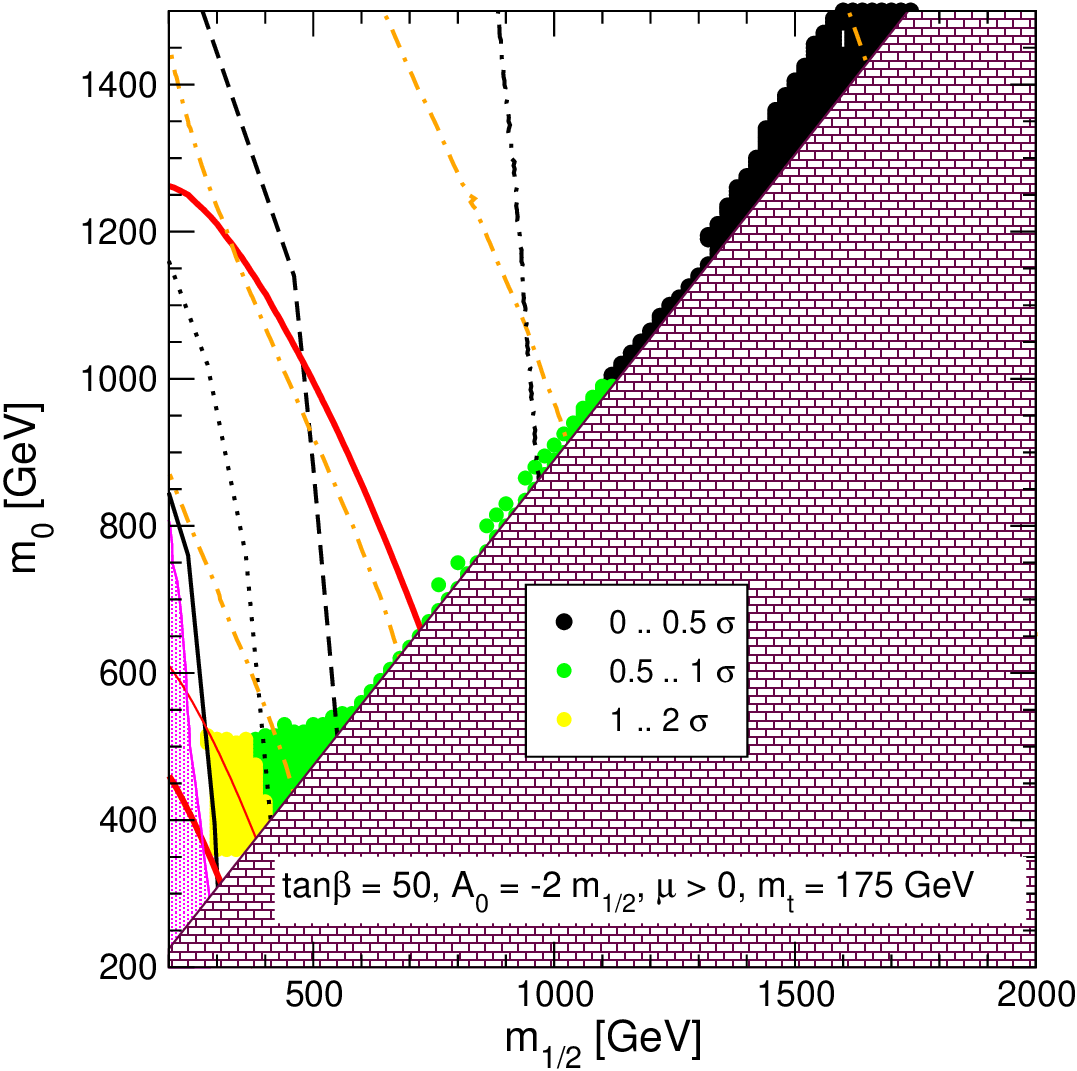,height=3.2in} \hfill
\end{minipage}

\caption{\it\small 
The deviations of \sibr{\eeZh}{\hbb} for the lightest $\cp$-even
CMSSM Higgs boson, normalized to the value in the SM with the same Higgs 
mass, are given in the
$(m_{1/2}, m_0)$ planes for $\mu > 0$, $\tb = 10, 50$ and $A_0 = 0$
(upper row), for $\mu > 0$, $\tb = 50$ and $A_0 = -2 m_{1/2}$ (lower
right) and for $\mu < 0$, $\tb = 35$ and $A_0 = m_{1/2}$ (lower left).
In all plots $\mt = 175 \gev$ has been used.
The diagonal red thick (thin) lines are the 
$\pm 2 - \si$ contours for $g_\mu - 2$: +56.3, +11.5 (+38.1, -4.7).
The near-vertical solid, dotted short-dashed, dash-dotted and
long-dashed (black) lines are the  
$\mh = 113, 115, 117, 120, 125 \gev$ contours. 
The lighter dot-dashed (orange) lines correspond to 
$\MA = 500, 700, 1000, 1500 \gev$.
The light shaded (pink) regions are excluded by 
$b \to s \ga$. 
The (brown) bricked
regions are excluded because the LSP is the charged $\Staue$ in these regions.
\label{fig:LCbb}
}
\vspace{-2em}
\end{figure}

Fig.~\ref{fig:LCbb} shows our results for \sibr{\eeZh}{\hbb} in the
$(m_{1/2}, m_0)$~plane. We have chosen the same representative set of
parameters for this and the products $\si \times \cB$ for all the other
channels. The upper left (right) plot shows the results for $\tb = 10
(50)$, $A_0 = 0$ and $\mu > 0$. The lower left plot shows $\tb = 35$, $A_0
= m_{1/2}$ and $\mu < 0$, and the lower right plot is for $\tb = 50$, $A_0
= -2 m_{1/2}$ and $\mu > 0$. We only show one plot for each channel for
$\mu < 0$, since this sign of $\mu$ is disfavored by the recent $g_\mu -
2$ measurement~\cite{g-2} as well as by the $\cB(b \to s \ga)$
constraint~\cite{bsgexp,bsgtheory}. The thick solid diagonal red lines
show the $\pm2-\si$ range of $g_\mu - 2$ for the standard value of the
discrepancy $\delta a_\mu = (33.9 \pm 11.2) \times 10^{-10}$, and the
thin red  lines
correspond to the more conservative estimate 
$\delta a_\mu = (16.7 \pm 10.7) \times 10^{-10}$. 
For $\mu > 0$, we have $2 - \si$ lower bounds on 
$(m_{1/2},
m_0)$ from both theory evaluations, while only the standard estimate also
results in upper bounds. The more conservative estimate also yields a thin
red line for $\mu < 0$, where the parameter space to the left is excluded
experimentally. The nearly vertical solid (dotted, short-dashed,
dot-dashed and long-dashed) black lines correspond to the 
contours~\cite{fh} $\mh = 113 \; (115, 117, 120, 125) \gev$, where the
latter one is only visible in the scenario with $\mu < 0$ for
large values of $m_{1/2}$. The
regions excluded by $\cB(b \to s \ga)$ are shown as the pink shaded areas,
that are more prominent in the $\mu < 0$ 
case~\footnote{In~\reffi{fig:LCbb}a, where the $b \to s \ga$ bound yields
much weaker constraints than the search limits on $\mh$ and the 
supersymmetric
particles, the region excluded by $b \to s \ga$ is not shown.}.
Finally, the large bricked
region in the lower right part of each plot corresponds to the region in
which the lightest $\Stau$ is the LSP, which is excluded because the LSP
cannot be charged. The colored area is that where the relic density of the
neutralino LSP is in the range $0.1 < \ohsq < 0.3$ preferred by cosmology.

In \reffi{fig:LCbb} and subsequent figures, we code with different
shadings domains of the CMSSM parameter space, consistent with the direct
and cosmological constraints mentioned above, where the CMSSM prediction
differs from the SM by different numbers of standard deviations $\sigma$,
as estimated on the basis of the precisions quoted in
Table~\ref{tab:channels}.

We see in \reffi{fig:LCbb} that the channel \sibr{\eeZh}{\hbb} exhibits
large deviations from the SM only for $\mu < 0$ and $m_{1/2} < 1000 \gev$,
a region that is excluded by $\cB(b \to s \ga)$ and disfavoured by $g_\mu
- 2$. For $\mu > 0$, deviations of 2 or 3 $\si$ are only observed for $\tb
= 10$ with very low $m_{1/2}$, otherwise the deviations are below the
$2-\si$ level. This means that, on the one hand, the LC will have no
problem in observing the lightest CMSSM Higgs boson in this channel. On
the other hand, it will not be easy to obtain additional indirect
information on the CMSSM Higgs sector by the precise measurement of this
channel; see, however, \refse{sec:MAreach}.

The interplay between the measurement of $\si \times \cB$ and the 
measurement of the Higgs-boson mass can be seen from the contour lines 
indicating different values of $\mh$. The compatibility of the
$\mh$ measurement with the results for $\si \times \cB$ (and with
possible information on the sparticle spectrum) is a stringent
consistency test of the CMSSM. For instance, a measurement of
$\mh \gsim 118 \gev$ in \reffi{fig:LCbb}a (upper left panel) would be
compatible only with values of \sibr{\eeZh}{\hbb} that differ from the SM
value by not more than one standard deviation. Observation of a
significantly larger deviation of \sibr{\eeZh}{\hbb} from the SM value
with $\mh \approx 118 \gev$
would disfavor an interpretation within the CMSSM for the parameters of
\reffi{fig:LCbb}a.

The results in the \sibr{\eeZh}{\htautau} channel, which are not shown
here, are similar in pattern to the \sibr{\eeZh}{\hbb} channel. The
deviations from the SM, however, are somewhat smaller in the $\htautau$
case.

\begin{figure}[!ht]

\begin{center}
\sibrfrac{\eeZh}{\hcc}
\end{center}

\vspace{-.1em}

\begin{minipage}{8in}
\hspace{4cm} (a) \hspace{7.2cm} (b) 
\end{minipage}

\vspace{0.2em}

\begin{minipage}{8in}
\epsfig{file=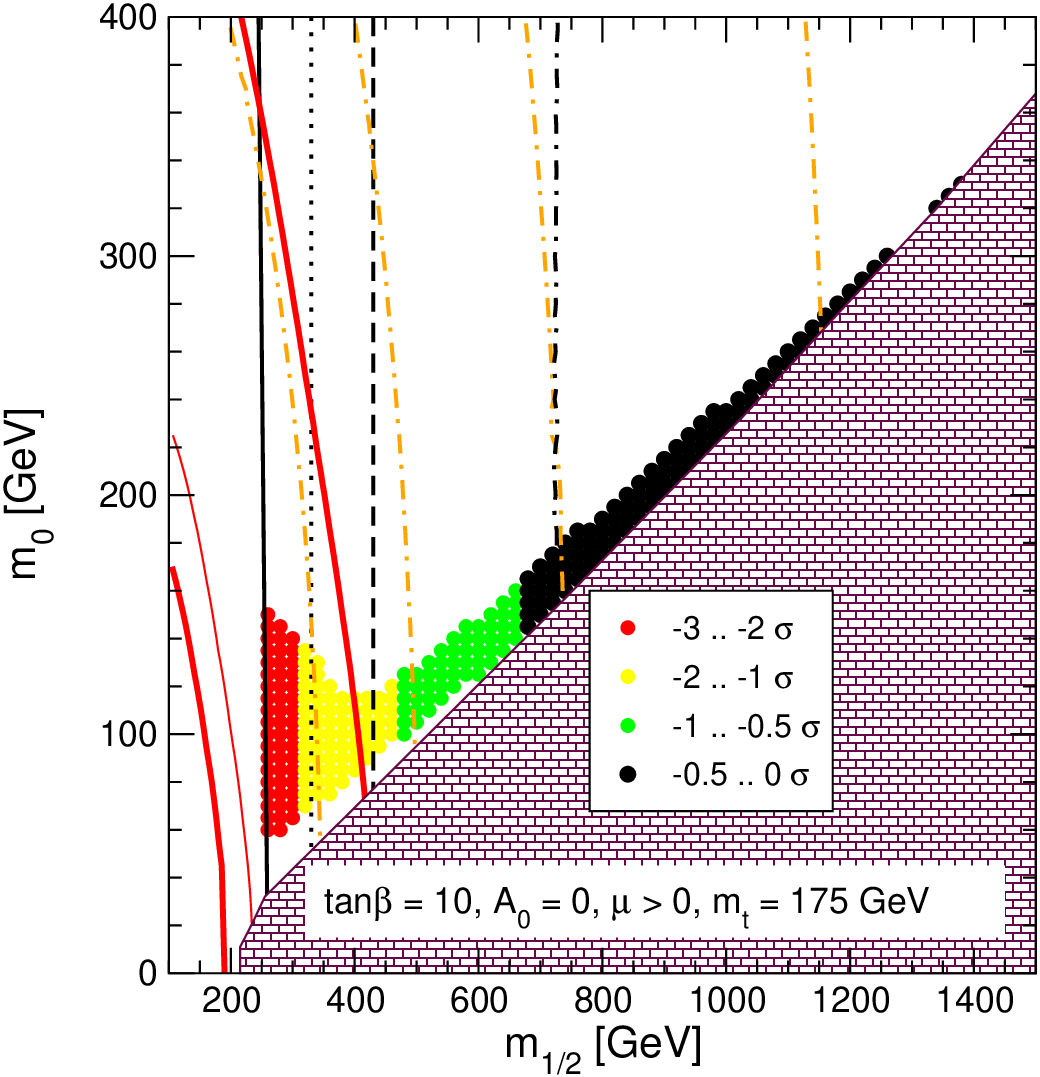,height=3.2in}
\epsfig{file=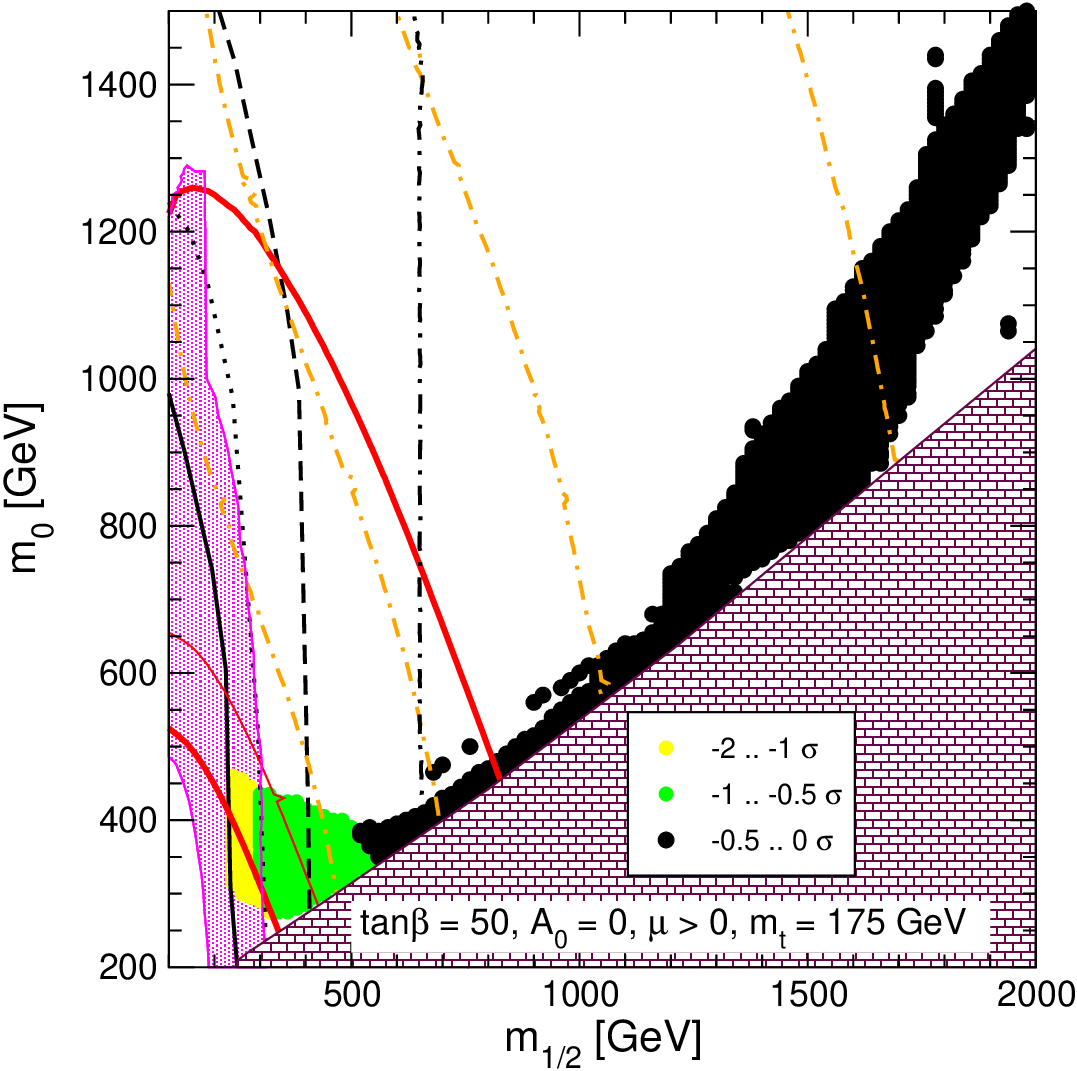,height=3.2in} \hfill
\end{minipage}

\vspace{2.5em}

\begin{minipage}{8in}
\hspace{4cm} (c) \hspace{7.2cm} (d) 
\end{minipage}

\vspace{.2em}

\begin{minipage}{8in}
\epsfig{file=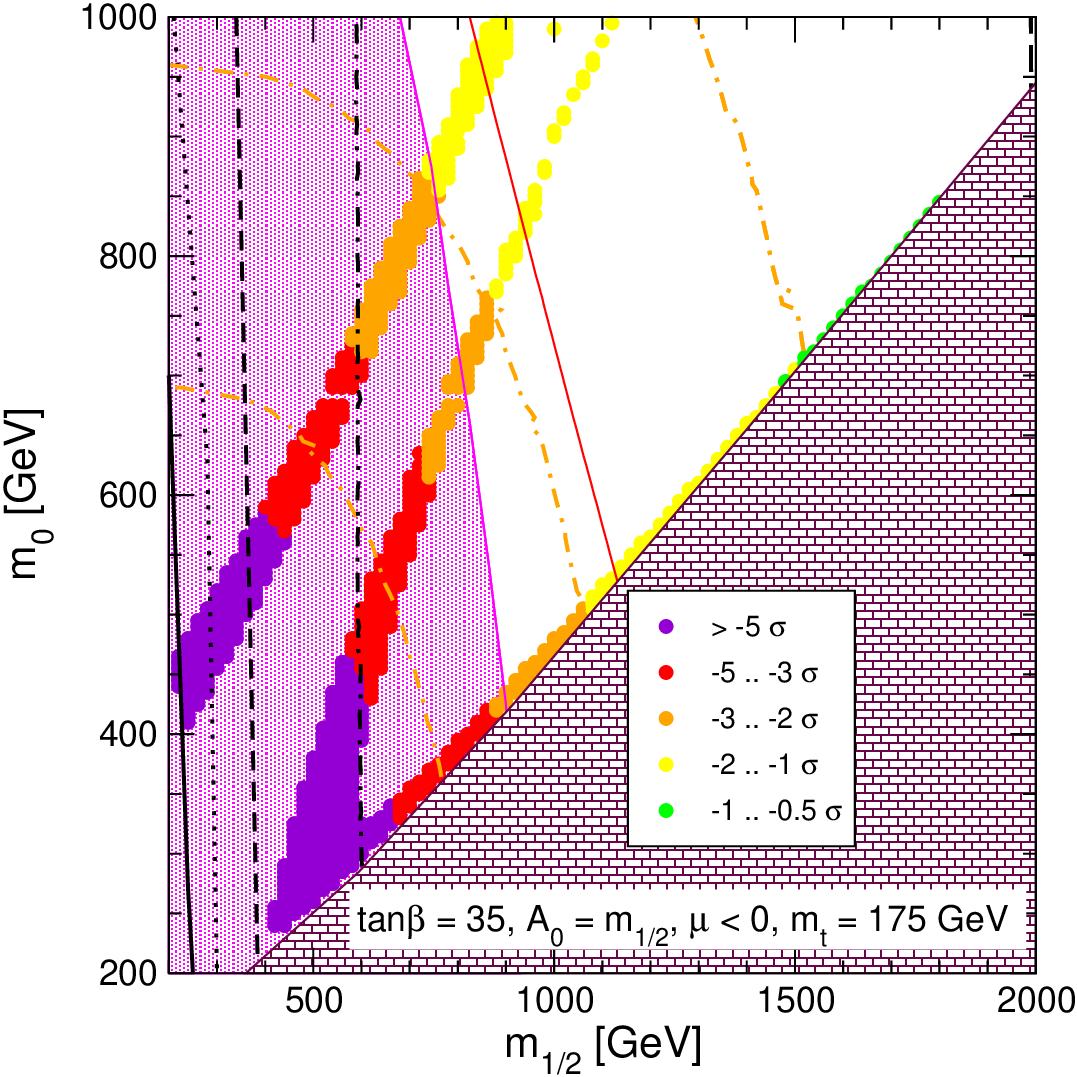,height=3.2in}
\epsfig{file=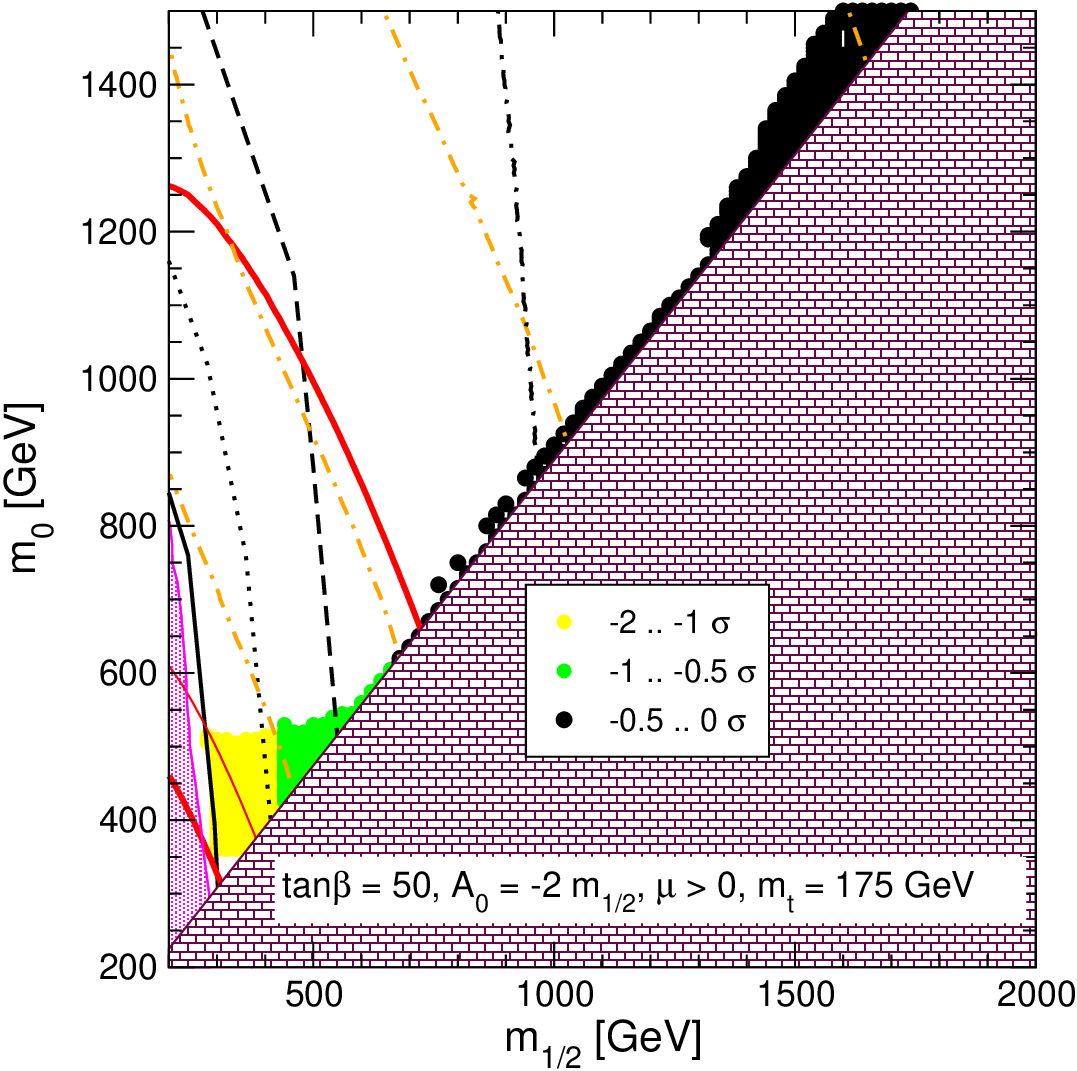,height=3.2in} \hfill
\end{minipage}

\vspace{1em}
\caption{\it\small 
The deviations of \sibr{\eeZh}{\hcc} for the lightest $\cp$-even
CMSSM Higgs boson, normalized to the value for a
SM Higgs boson with the same mass, are given in the
$(m_{1/2}, m_0)$ planes for $\mu > 0$, $\tb = 10, 50$ and $A_0 = 0$
(upper row), for $\mu > 0$, $\tb = 50$ and $A_0 = -2 m_{1/2}$ (lower
right) and for $\mu < 0$, $\tb = 35$ and $A_0 = m_{1/2}$ (lower left).
In all plots $\mt = 175 \gev$ has been used.
The contours and shadings are the same as in Fig.~\ref{fig:LCbb}.
\label{fig:LCcc}
}
\end{figure}

\reffi{fig:LCcc} shows the channel \sibr{\eeZh}{\hcc}. Here the situation
is different from that in the $\hbb$ decay channel, as the CMSSM result is
always somewhat smaller than the corresponding SM result. This is due to
the enlarged value of $\cB(\hbb)$, see \reffi{fig:LCbb}, which reduces the
branching ratio for the $\hcc$ mode. The absolute sizes of the
deviations are similar to those in the $\hbb$ case. As before, the largest
deviations again occur for negative $\mu$ in the experimentally excluded
part of the parameter space. The largest deviations in the allowed part of
parameter space are for $\tb = 10$ and small $m_{1/2}$, where 2-
to 3-$\sigma$ deviations may be attained.

\begin{figure}[!ht]

\begin{center}
\sibrfrac{\eeZh}{\hWW}
\end{center}

\vspace{-.1em}

\begin{minipage}{8in}
\hspace{4cm} (a) \hspace{7.2cm} (b) 
\end{minipage}

\vspace{.2em}

\begin{minipage}{8in}
\epsfig{file=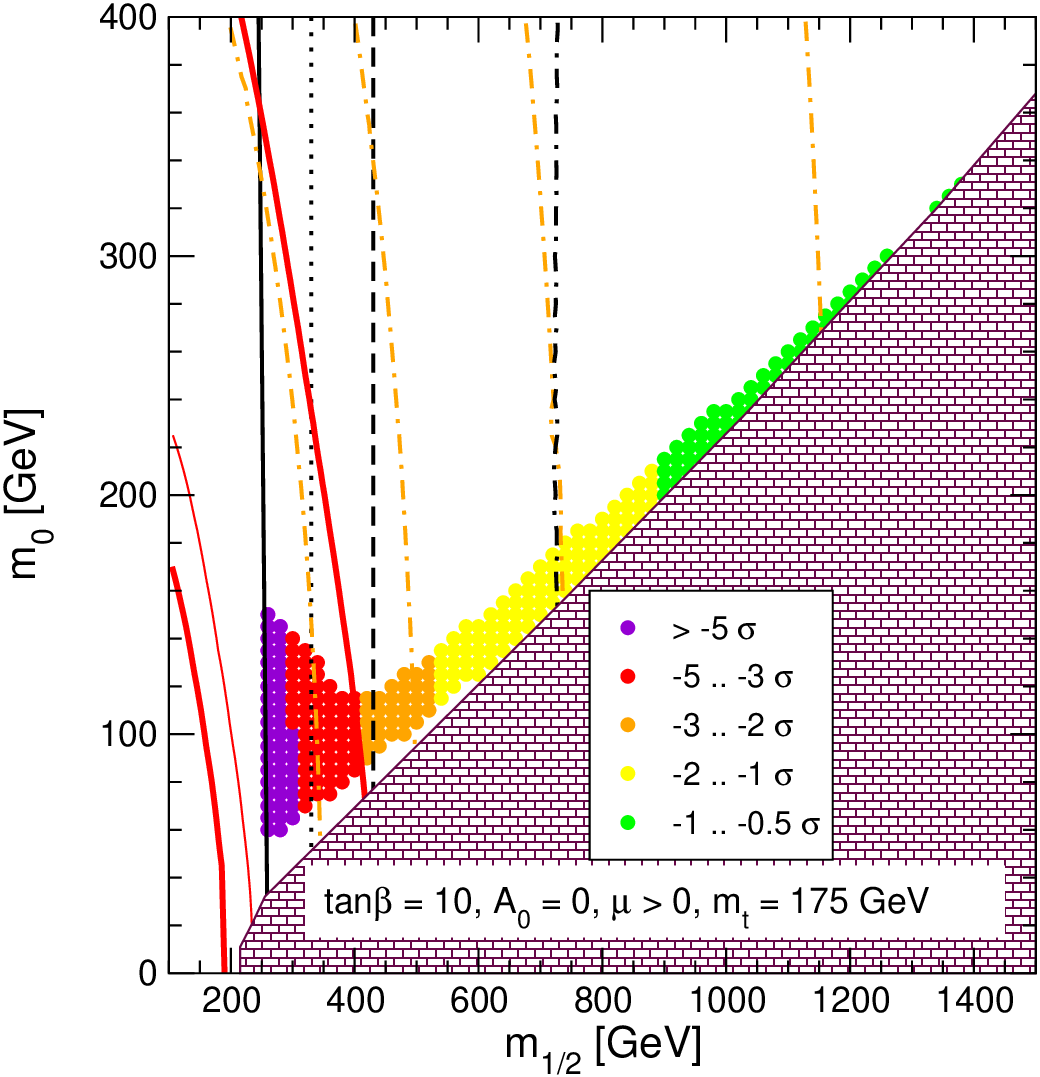,height=3.2in}
\epsfig{file=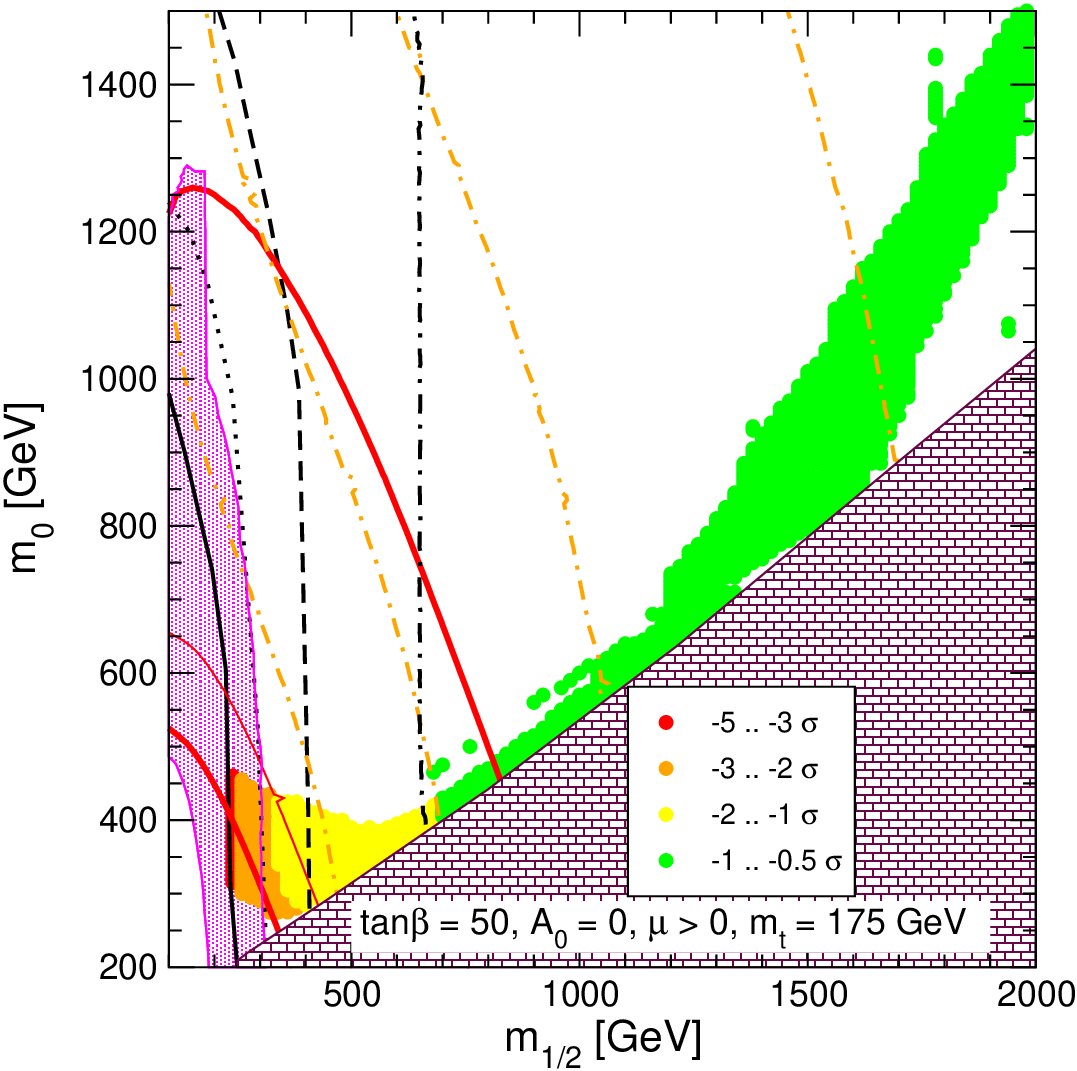,height=3.2in} \hfill
\end{minipage}

\vspace{2.5em}

\begin{minipage}{8in}
\hspace{4cm} (c) \hspace{7.2cm} (d) 
\end{minipage}

\vspace{.2em}

\begin{minipage}{8in}
\epsfig{file=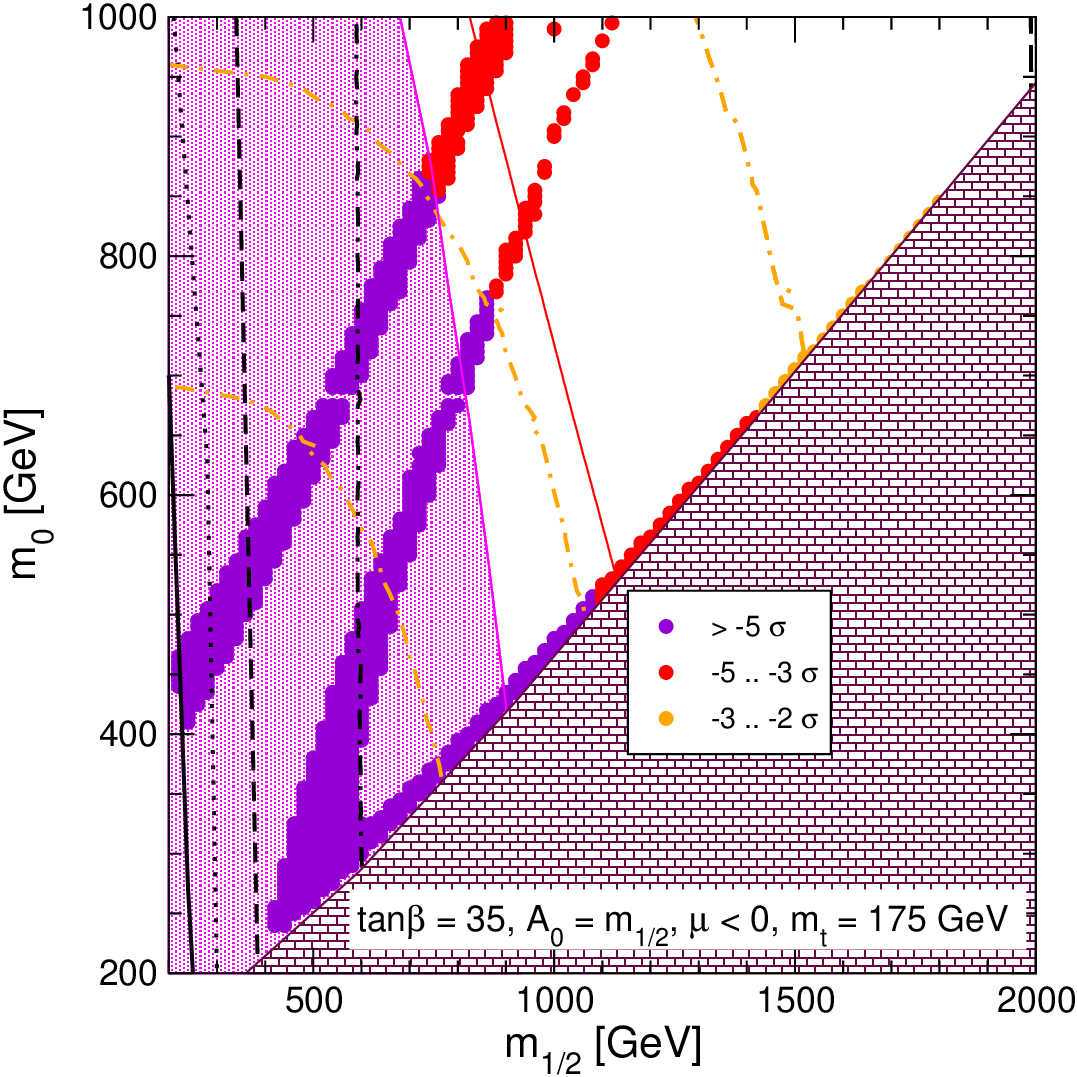,height=3.2in}
\epsfig{file=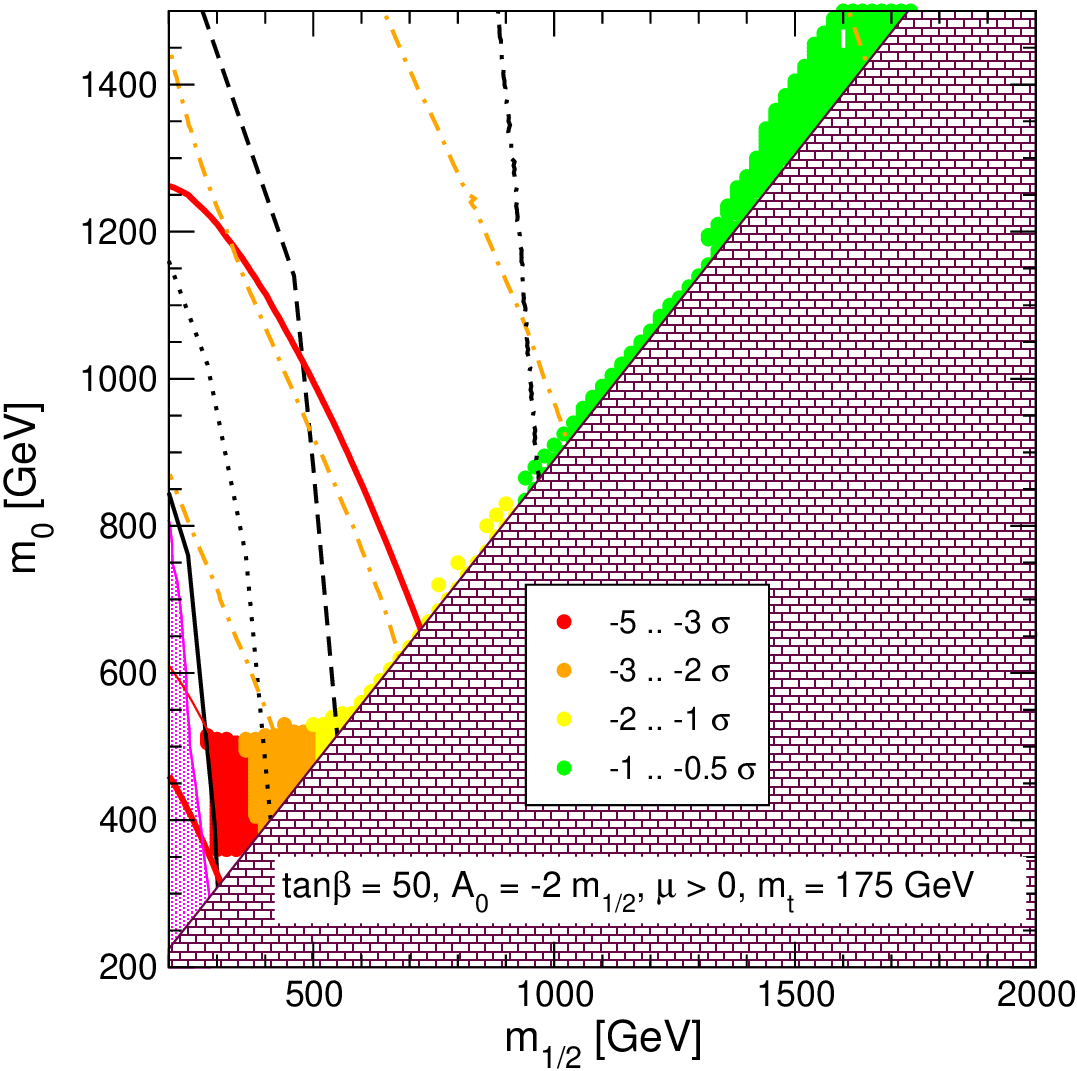,height=3.2in} \hfill
\end{minipage}

\vspace{1em}
\caption{\it\small 
The deviations of \sibr{\eeZh}{\hWW} for the lightest $\cp$-even
CMSSM Higgs boson, normalized to the value for a 
SM Higgs boson with the same mass, are given in the
$(m_{1/2}, m_0)$ planes for $\mu > 0$, $\tb = 10, 50$ and $A_0 = 0$
(upper row), for $\mu > 0$, $\tb = 50$ and $A_0 = -2 m_{1/2}$ (lower
right) and for $\mu < 0$, $\tb = 35$ and $A_0 = m_{1/2}$ (lower left).
In all plots $\mt = 175 \gev$ has been used.
The contours and shadings are the same as in 
Fig.~\ref{fig:LCbb}.
\label{fig:LCWW}
}
\end{figure}

The LC survey is completed by \reffi{fig:LCWW}, in which the
\sibr{\eeZh}{\hWW} channel is shown. Here the CMSSM signal is always
smaller than the corresponding SM result, with the enhancement of $\cB
(\hbb)$ again playing a role. In addition, the decay $\hWW$ is always
suppressed in the MSSM, because of the additional coupling factor 
$\SQba \lsim 1$, though this suppression has only a marginal effect
in the CMSSM parameter space.
The largest deviations again occur for $\mu < 0$, but deviations of
$5\si$ or more can be observed also for $\mu > 0$. 
While such a suppression should not endanger the observability of this
channel, the sizable deviations from the SM prediction provide a
sensitive consistency test of the CMSSM and allow one to obtain 
valuable indirect information on the Higgs boson sector, as shown in
more detail in \refse{sec:MAreach}.


\subsection{Observables at a $\ga \ga$ Collider}

\begin{figure}[!ht]
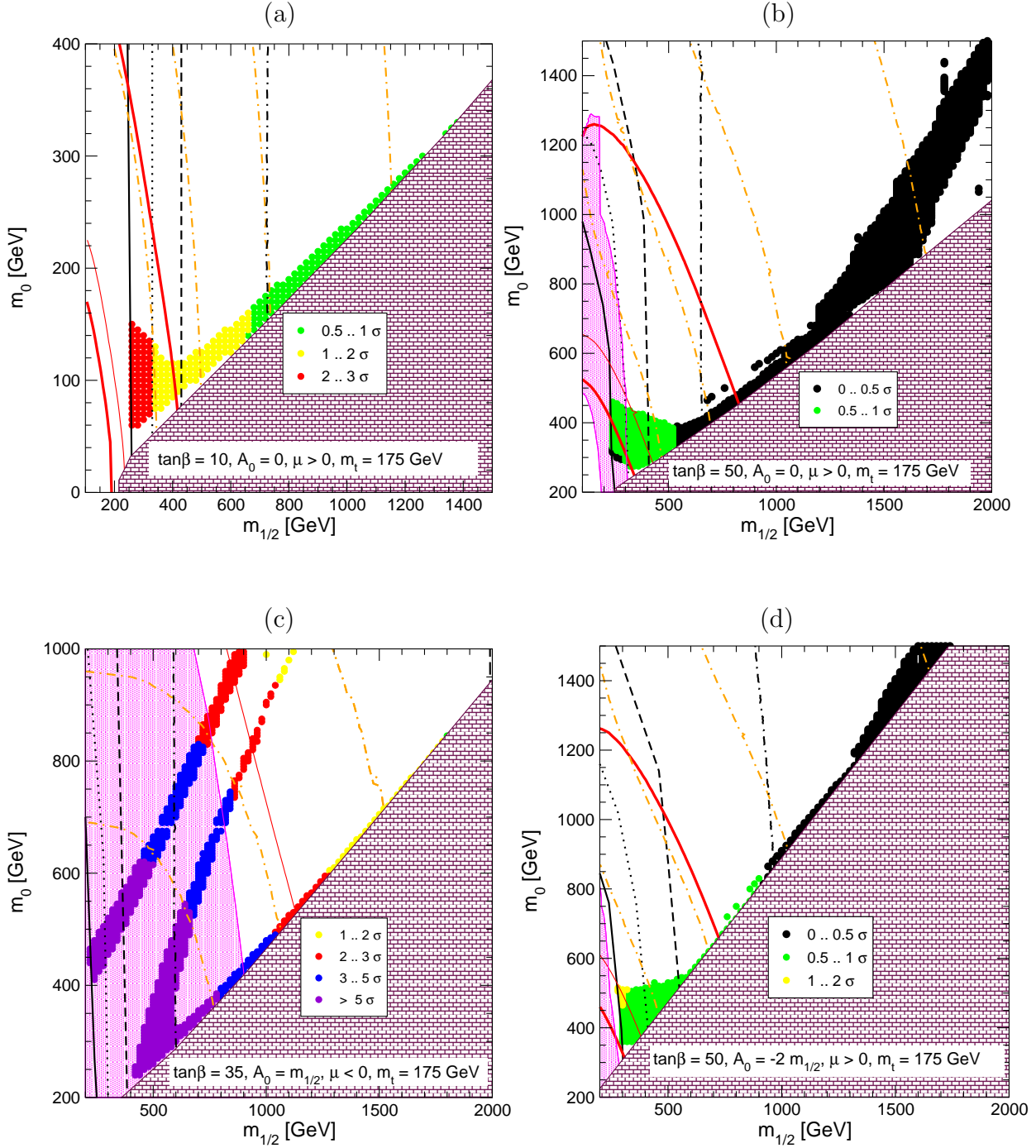


\begin{center}
\sibrfrac{\gagah}{\hbb}
\end{center}

\vspace{-.1em}

\begin{minipage}{8in}
\hspace{4cm} (a) \hspace{7.2cm} (b) 
\end{minipage}

\vspace{.2em}

\begin{minipage}{8in}
\epsfig{file=EHOW22e.03.cl.eps,height=3.2in}
\epsfig{file=EHOW22e.09.cl.eps,height=3.2in} \hfill
\end{minipage}

\vspace{2.5em}

\begin{minipage}{8in}
\hspace{4cm} (c) \hspace{7.2cm} (d) 
\end{minipage}

\vspace{.2em}

\begin{minipage}{8in}
\epsfig{file=EHOW22e.24.cl.eps,height=3.2in}
\epsfig{file=EHOW22e.20.cl.eps,height=3.2in} \hfill
\end{minipage}

\vspace{1em}
\caption{\it\small 
The deviations of \sibr{\gagah}{\hbb} for the lightest $\cp$-even
CMSSM Higgs boson, normalized to the value in the SM with the same 
Higgs mass, are given in the
$(m_{1/2}, m_0)$ planes for $\mu > 0$, $\tb = 10, 50$ and $A_0 = 0$
(upper row), for $\mu > 0$, $\tb = 50$ and $A_0 = -2 m_{1/2}$ (lower
right) and for $\mu < 0$, $\tb = 35$ and $A_0 = m_{1/2}$ (lower left).
The contours and shadings are the same as in Fig.~\ref{fig:LCbb}.
\label{fig:gaCbb}
}
\end{figure}

We now turn to the channels accessible at a \gaC, which might become
available some time after the construction of a LC. In \reffi{fig:gaCbb}
we first show results for the main Higgs decay channel,
\sibr{\gagah}{\hbb}. This looks quite similar to the corresponding LC
results, and the same is true for the \sibr{\gagah}{\hWW} channel, which
is not shown here.

\begin{figure}[!ht]

\begin{center}
\sibrfrac{\gagah}{\hgaga}
\end{center}


\begin{minipage}{8in}
\hspace{4cm} (a) \hspace{7.2cm} (b) 
\end{minipage}

\vspace{.2em}

\begin{minipage}{8in}
\epsfig{file=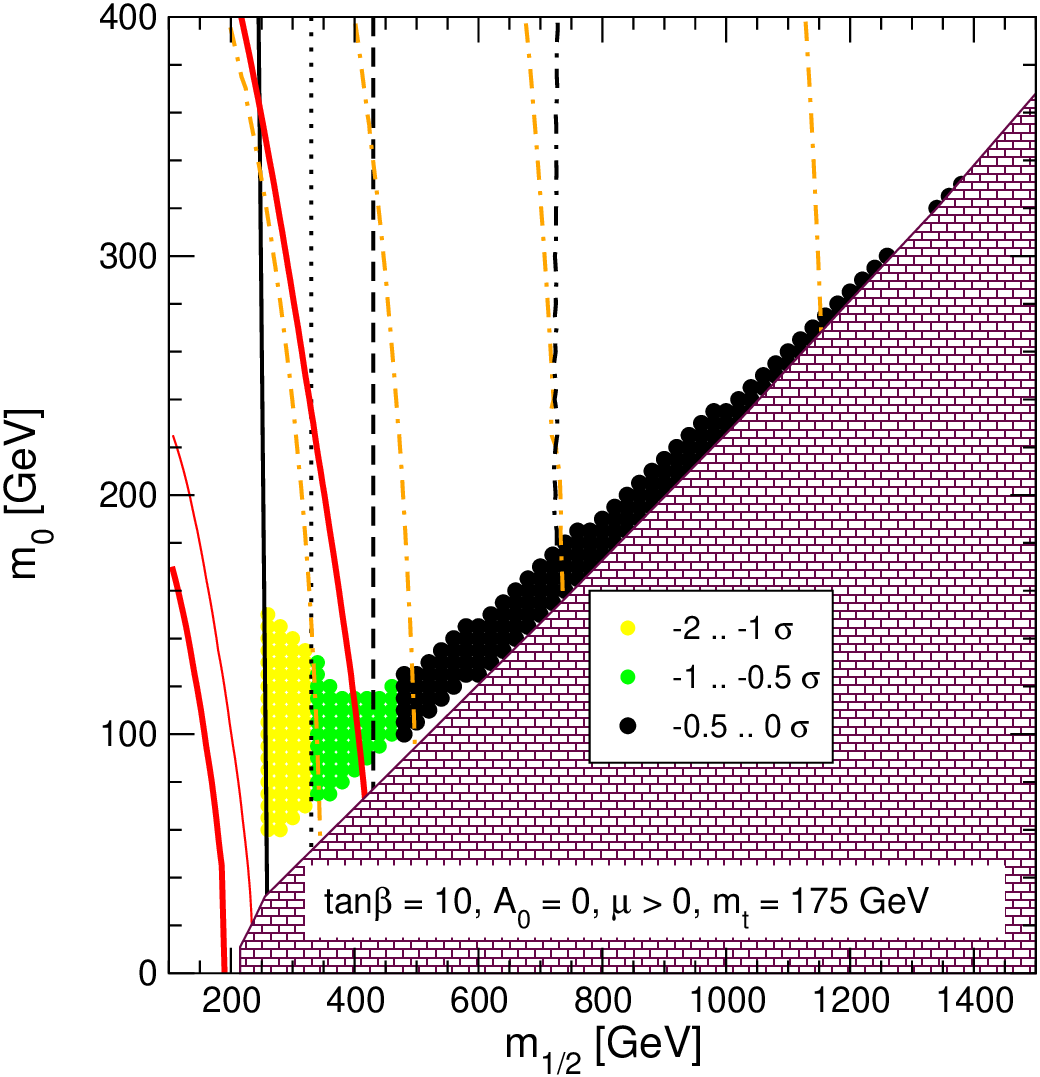,height=3.2in}
\epsfig{file=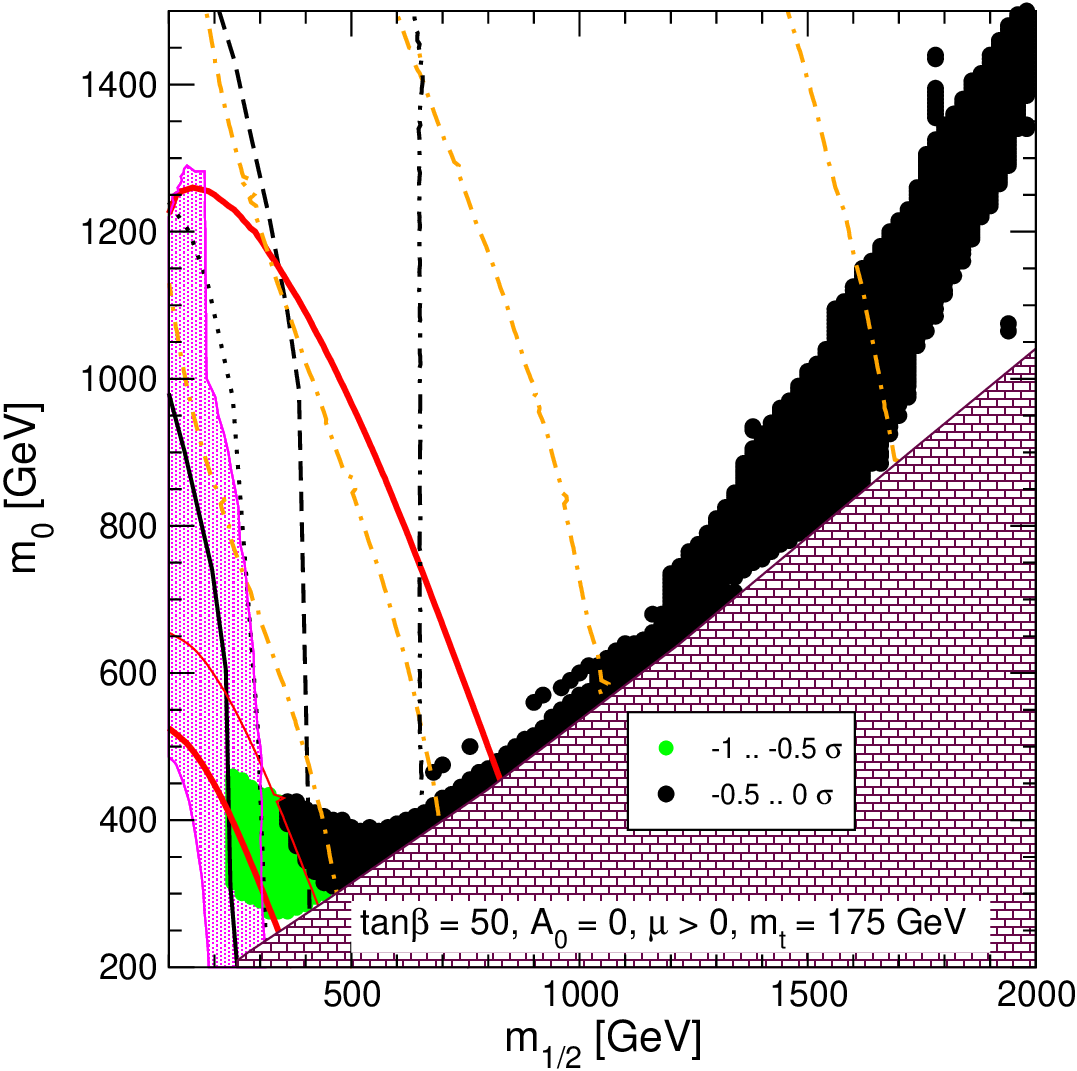,height=3.2in} \hfill
\end{minipage}

\vspace{2.5em}

\begin{minipage}{8in}
\hspace{4cm} (c) \hspace{7.2cm} (d) 
\end{minipage}

\vspace{.2em}

\begin{minipage}{8in}
\epsfig{file=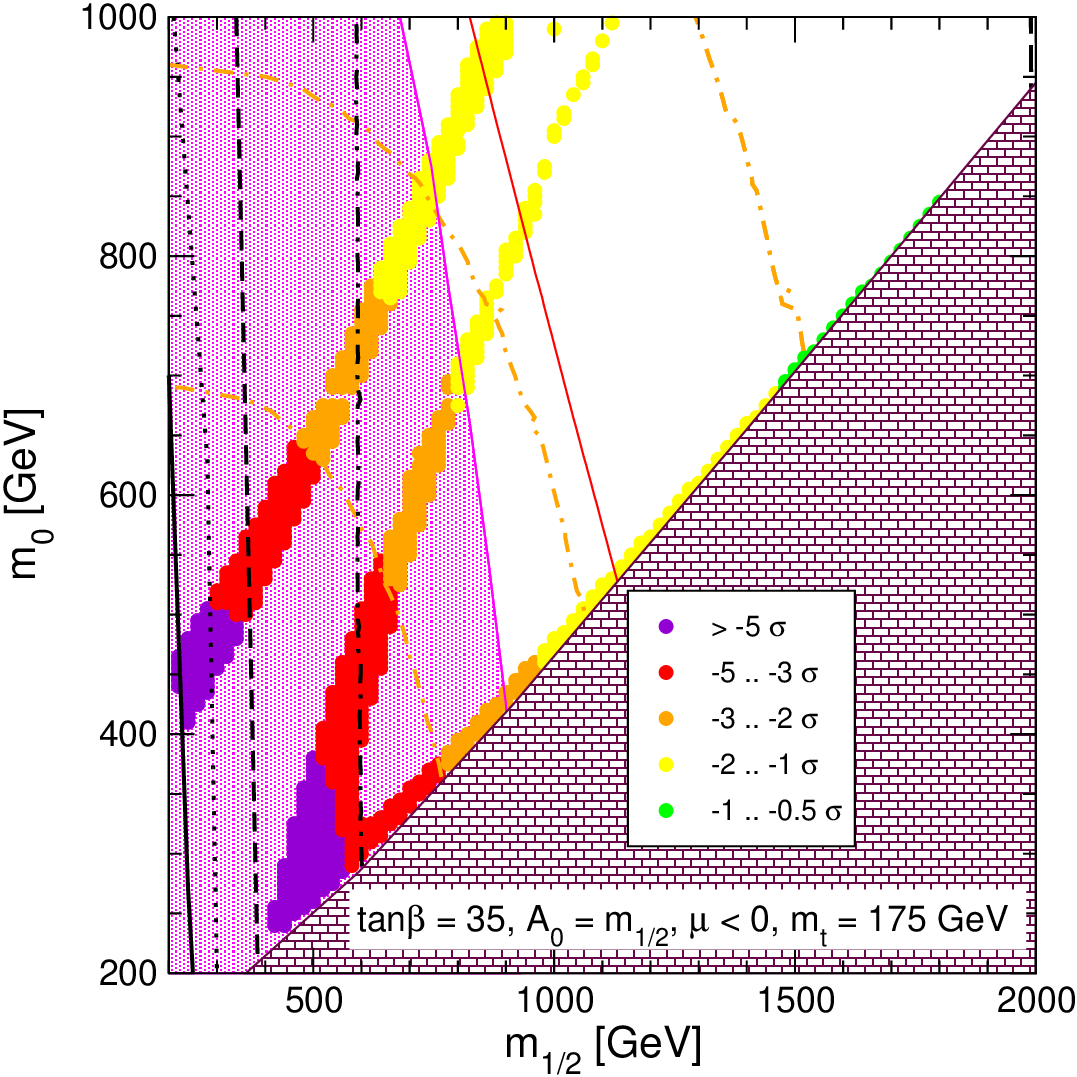,height=3.2in}
\epsfig{file=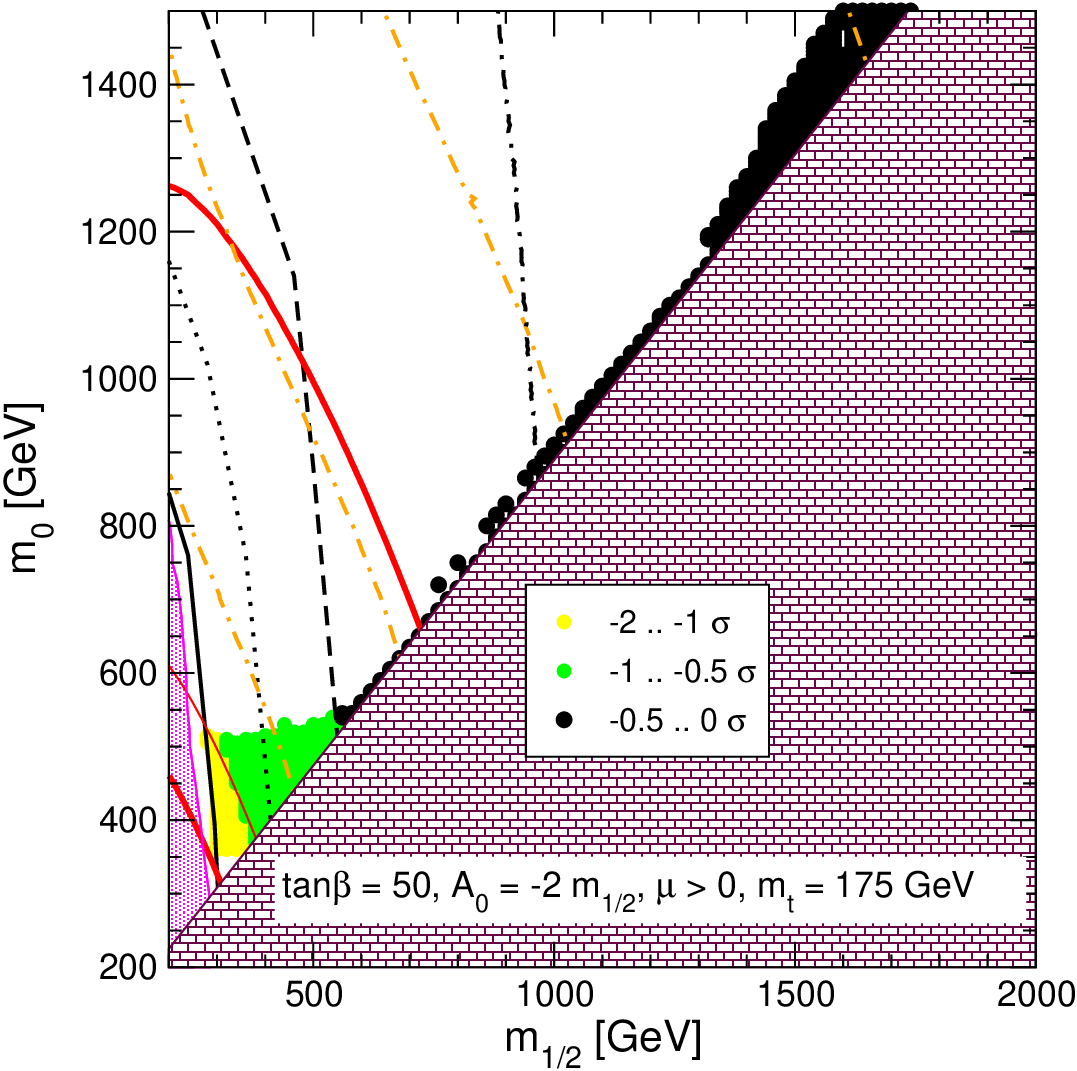,height=3.2in} \hfill
\end{minipage}

\caption{\it\small 
The deviations of \sibr{\gagah}{\hgaga} for the lightest $\cp$-even
CMSSM Higgs boson, normalized to the value for a SM Higgs boson with
the mass, are given in the
$(m_{1/2}, m_0)$ planes for $\mu > 0$, $\tb = 10, 50$ and $A_0 = 0$
(upper row), for $\mu > 0$, $\tb = 50$ and $A_0 = -2 m_{1/2}$ (lower
right) and for $\mu < 0$, $\tb = 35$ and $A_0 = m_{1/2}$ (lower left).
In all plots $\mt = 175 \gev$ has been used.
The contours and shadings are the same as in Fig.~\ref{fig:LCbb}.
\label{fig:gaCgaga}
}
\end{figure}

We show in \reffi{fig:gaCgaga} results for the additional observable at
the \gaC, namely \sibr{\gagah}{\hgaga}. This channel can be isolated from
the background using the feature that the signal, contrary to the
background, peaks sharply in the forward region~\cite{gaC1,gaC2}. However,
the precision expected is only at the 11\% level, so that no large
deviations will be observable. As seen in Fig.~\ref{fig:gaCgaga}, for $\mu
> 0$ they always stay below the $2-\si$ level, and only in the excluded
region with $\mu < 0$ do larger deviations occur.

\subsection{Observables at a $\mu^+ \mu^-$ Collider}

\begin{figure}[!ht]

\begin{center}
\sibrfrac{\mumuh}{\hbb}
\end{center}

\vspace{-1em}

\begin{minipage}{8in}
\hspace{4cm} (a) \hspace{7.2cm} (b) 
\end{minipage}

\vspace{.2em}

\begin{minipage}{8in}
\epsfig{file=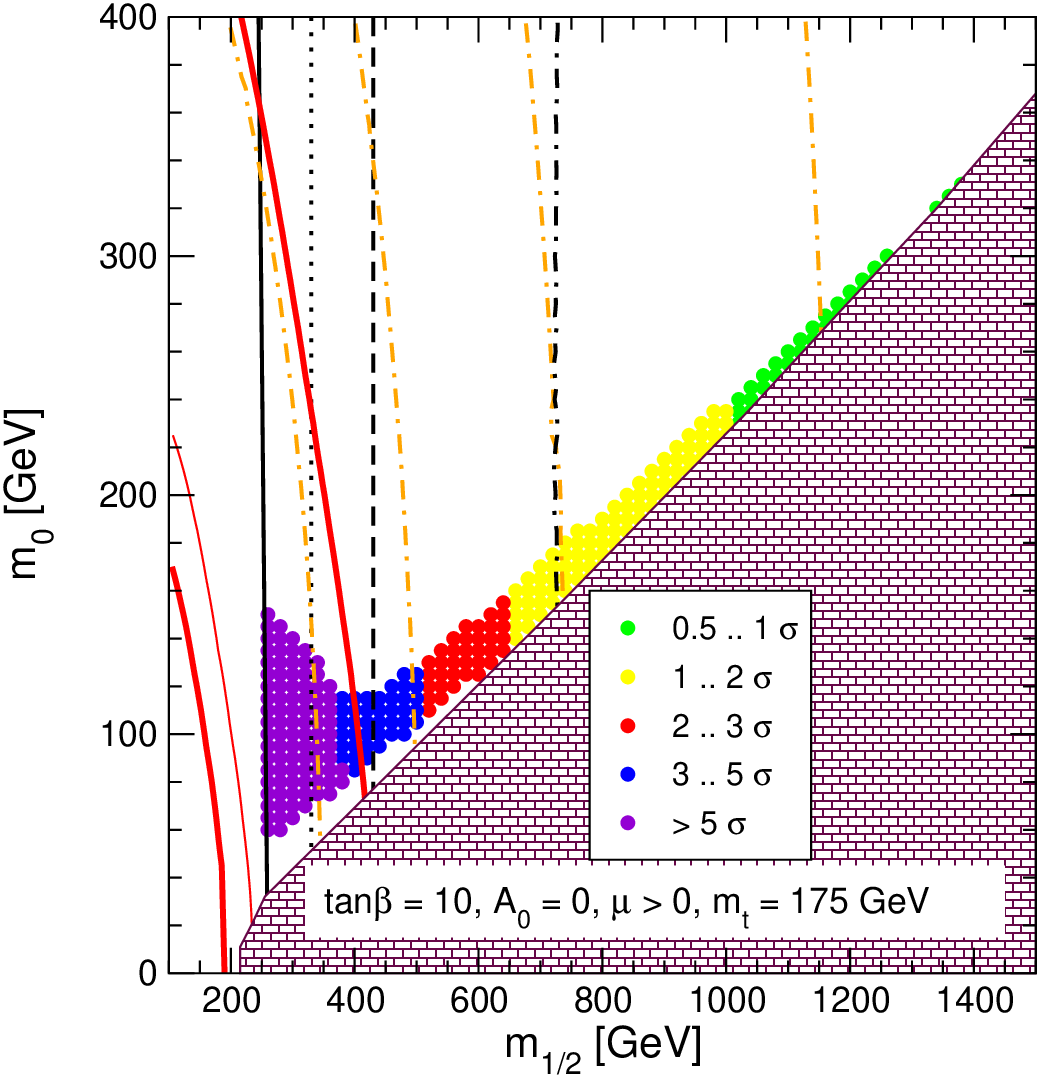,height=3.2in}
\epsfig{file=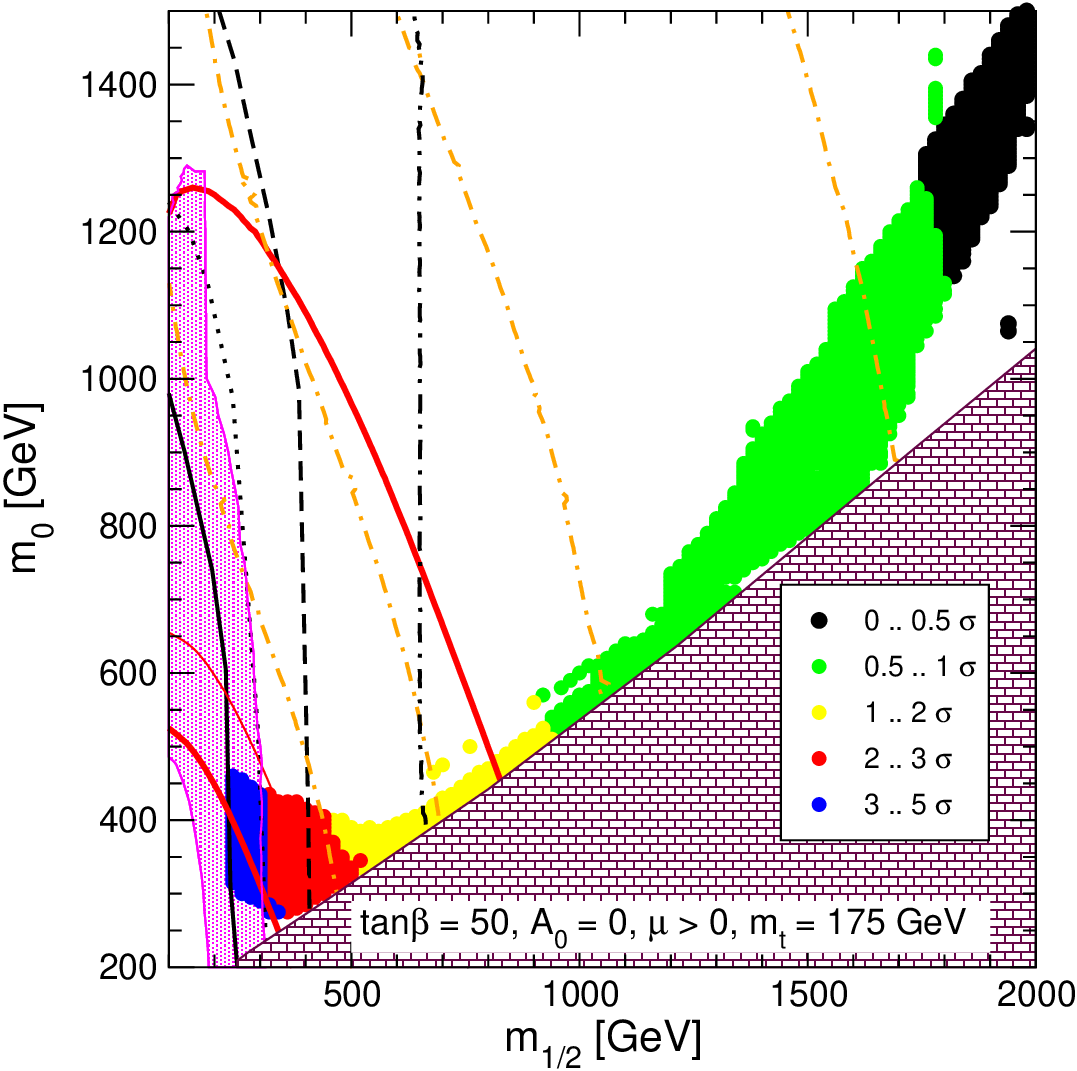,height=3.2in} \hfill
\end{minipage}

\vspace{2.5em}

\begin{minipage}{8in}
\hspace{4cm} (c) \hspace{7.2cm} (d) 
\end{minipage}

\vspace{.2em}

\begin{minipage}{8in}
\epsfig{file=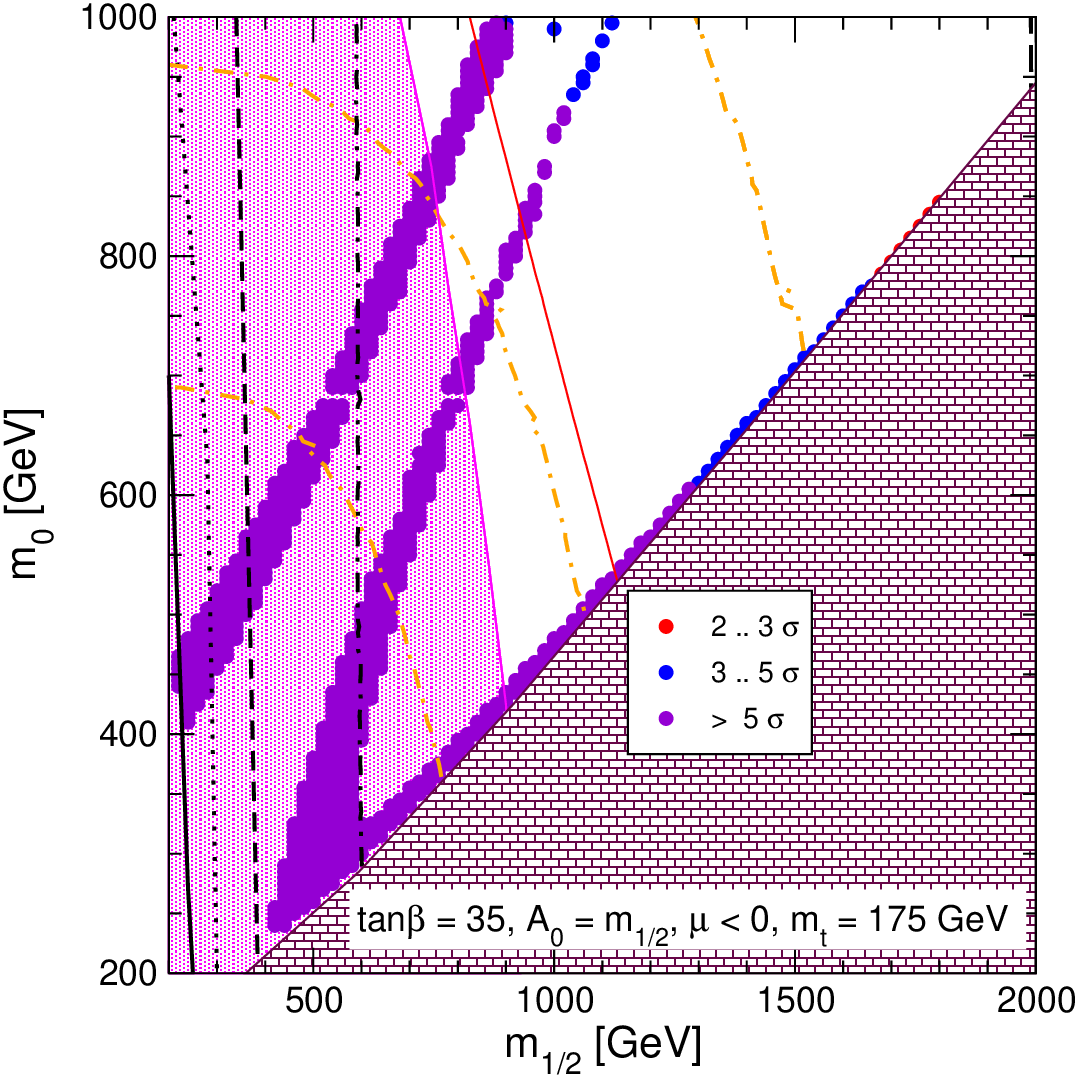,height=3.2in}
\epsfig{file=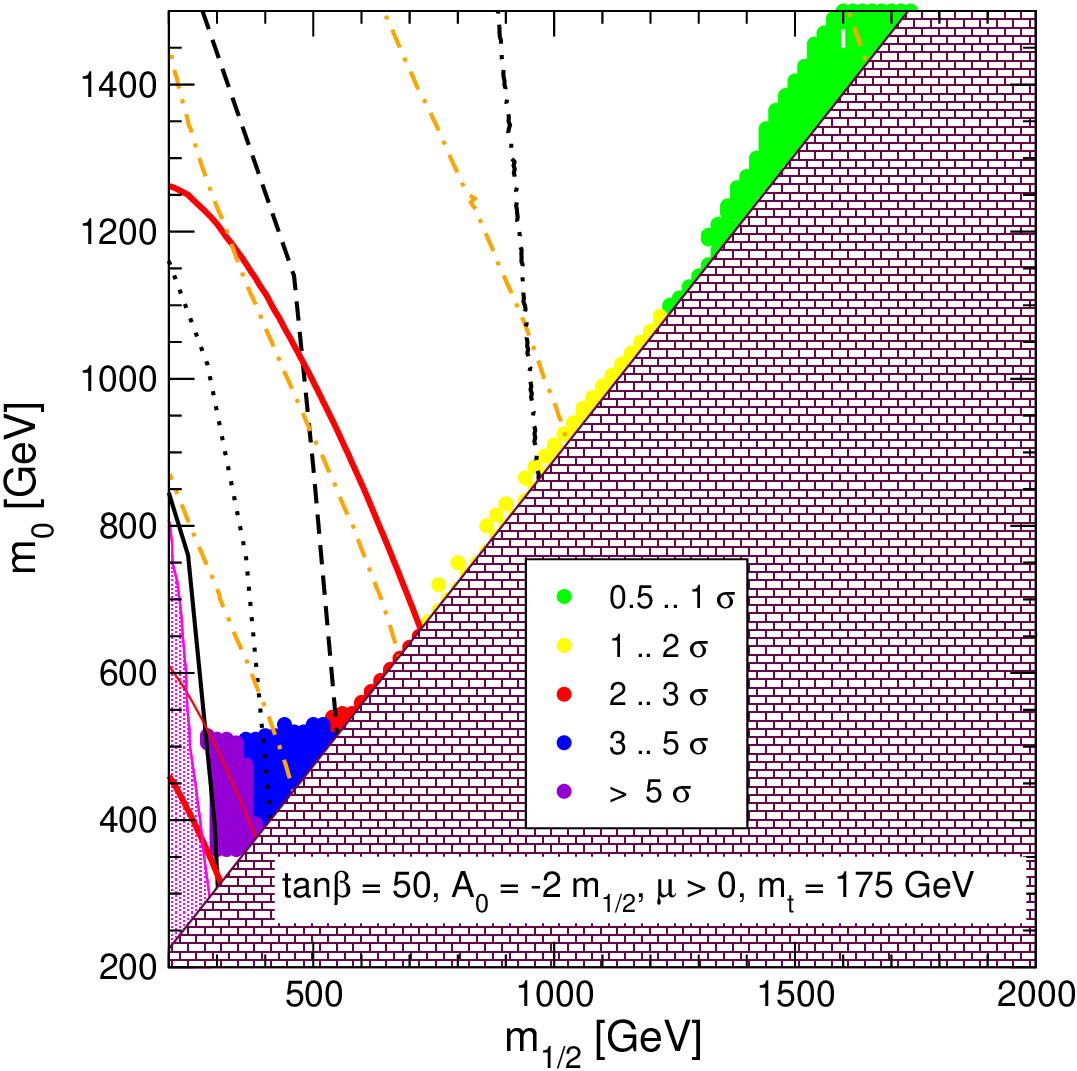,height=3.2in} \hfill
\end{minipage}

\vspace{1em}
\caption{\it\small 
The deviations of \sibr{\mumuh}{\hbb} for the lightest $\cp$-even
CMSSM Higgs boson, normalized to the value in the SM with the same 
Higgs mass, are given in the
$(m_{1/2}, m_0)$ planes for $\mu > 0$, $\tb = 10, 50$ and $A_0 = 0$
(upper row), for $\mu > 0$, $\tb = 50$ and $A_0 = -2 m_{1/2}$ (lower
right) and for $\mu < 0$, $\tb = 35$ and $A_0 = m_{1/2}$ (lower left).
In all plots $\mt = 175 \gev$ has been used.
The contours and shadings are the same as in Fig.~\ref{fig:LCbb}.
\label{fig:muCbb}
}
\end{figure}

This survey is completed with the main channel at the \muC, namely the
channel \sibr{\mumuh}{\hbb} shown in \reffi{fig:muCbb}. Due to the similar
coupling structures in the Higgs production channel, $\mumuh$, and in the
Higgs decay channel, $\hbb$, the deviations can be relatively large,
potentially exceeding $5\si$ even for positive $\mu$.  Therefore, in
principle, it should be possible to extract more indirect information
about the CMSSM Higgs boson sector. However, we expect that any \muC\ will
come into operation only after the other colliders considered above. 
Thus, a large variety of direct and indirect results should already be
available when the \muC\ starts running, making it rather difficult to
speculate about the impact of the prospective \muC\ precision
measurements. The \muC\ has in particular the potential for measuring 
$\mh$ with a spectacular precision~\cite{muC1}, which however needs to
be confronted with the parametric and higher-order uncertainties in the
Higgs-mass prediction. For investigating the MSSM Higgs sector, a
higher-energy \muC\ capable of producing directly the $\cp$-odd and -even
CMSSM Higgs bosons $A, H$ might also become an interesting option, but
studying the potential of such a machine
lies beyond the scope of this paper.


\section{Testing the CMSSM Universality Assumption and Indirect
Constraints on $\MA$}
\label{sec:MAreach}

The MSSM Higgs sector can be characterized at lowest order by the values 
of $\tb$ and the mass of the $\cp$-odd Higgs boson, $\MA$. As already 
mentioned, at the LHC there is a substantial part of parameter space where 
the heavy Higgs bosons cannot be observed, the so-called `wedge
region'~\cite{atlastdr,LHHiggsProcs2001}. On the other hand,  $\MA$
might well be too large for direct observation at the LC: $\MA \gsim
\sqrt{s}/2$.  In this case one would have to rely on indirect methods to
constrain the possible values of $\MA$.
Exploratory studies in some (favorable)
scenarios can be found in~\cite{hff,MAindirect}, also including
information from GigaZ~\cite{gigaz}.

In the CMSSM, $\MA$ is fixed by the electroweak vacuum conditions in terms
of $\tb$, $m_{1/2}$, $m_0$ and $A_0$. However, in a more general scenario 
like a non-universal Higgs model (NUHM)~\cite{CMSSMuni,EFOS} the CMSSM 
assumption that the soft supersymmetry-breaking masses 
for the Higgs multiplets are the same as those for squarks and sleptons
may be relaxed. In this case 
the values of $\MA$ and $\mu$ become independent parameters, though their 
ranges are restricted by various theoretical and phenomenological 
constraints. 
Accordingly, even if experimental results on the spectrum of
supersymmetric particles turn out to be compatible with the predictions
of the CMSSM scenario, confronting the CMSSM prediction for $\MA$ with 
direct or indirect information on this parameter provides a non-trivial 
test of the model. 

In the decoupling limit, $\MA \gg \MZ$, the couplings of the light
$\cp$-even Higgs boson of the MSSM become equal to those of the SM
Higgs boson. Thus, the observation of deviations in the production and 
decay of the lightest $\cp$-even Higgs boson of the MSSM from the SM 
values would allow one to set an upper (or lower) limit on the mass of
$\MA$. 

In order to facilitate the analysis of
indirect sensitivities to $\MA$ at different colliders in this context,
we have also shown in \reffi{fig:LCbb} - \ref{fig:muCbb} the
contour lines for $\MA = 500, 700, 1000, 1500 \gev$, as light (orange)
dot-dashed lines from left to right. Sometimes the $1500 \gev$ line is
missing; it would appear at larger $m_{1/2}$. For given values of $A_0$,
$\tb$ and the sign of $\mu$, these contour lines indicate that a
measurement of $\si \times \cB$ in a certain channel translates into an
allowed interval of $\MA$ values. 

The contour lines indicating different
values of $\MA$ within the CMSSM can also be interpreted within the
NUHM. For the same values of the parameters $m_0$, $m_{1/2}$,
$A_0$, $\tb$ and $\mu$ in the NUHM as in the CMSSM (see below for a 
discussion
of the $\mu$-dependence in the NUHM), decreasing the value of
$\MA$ within the NUHM compared to the value in the CMSSM will increase
the deviation of $\si \times \cB$ from the SM prediction and vice versa.
In this way measurements of $\si \times \cB$ can be used to establish an
upper bound on $\MA$ within the NUHM.

We first focus on the LC.  For $\tb = 10$, the channel \sibr{\eeZh}{\hWW}
offers the best prospects. An observation of a deviation of more than 3
(2, 1) $\si$ can be interpreted as an upper limit on $\MA$ of $\sim 600
\; (750, 1200) \gev$. We note that the decays $\hbb$ and $\hcc$ also show
some sensitivity for this parameter set. However, the situation is
somewhat worse when $\tb = 50$. For large and negative $A_0 = -2
m_{1/2}$, limits of $\MA \lsim 550, 900 \gev$ can be set at the $2-,
1-\si$ level, but the sensitivity decreases with increasing $A_0$. In
particular, for $A_0 = +m_{1/2}$ only very small deviations from the SM
predictions occur, and there is hardly any capability of setting an upper
bound on $\MA$.

\begin{figure}[!ht]

\mbox{}
\vspace{-1em}
\begin{center}
\sibrfrac{\eeZh}{\hbb}
\end{center}

\vspace{-.1em}

\begin{minipage}{8in}
\hspace{4cm} (a) \hspace{7.2cm} (b) 
\end{minipage}

\vspace{.2em}

\begin{minipage}{8in}
\epsfig{file=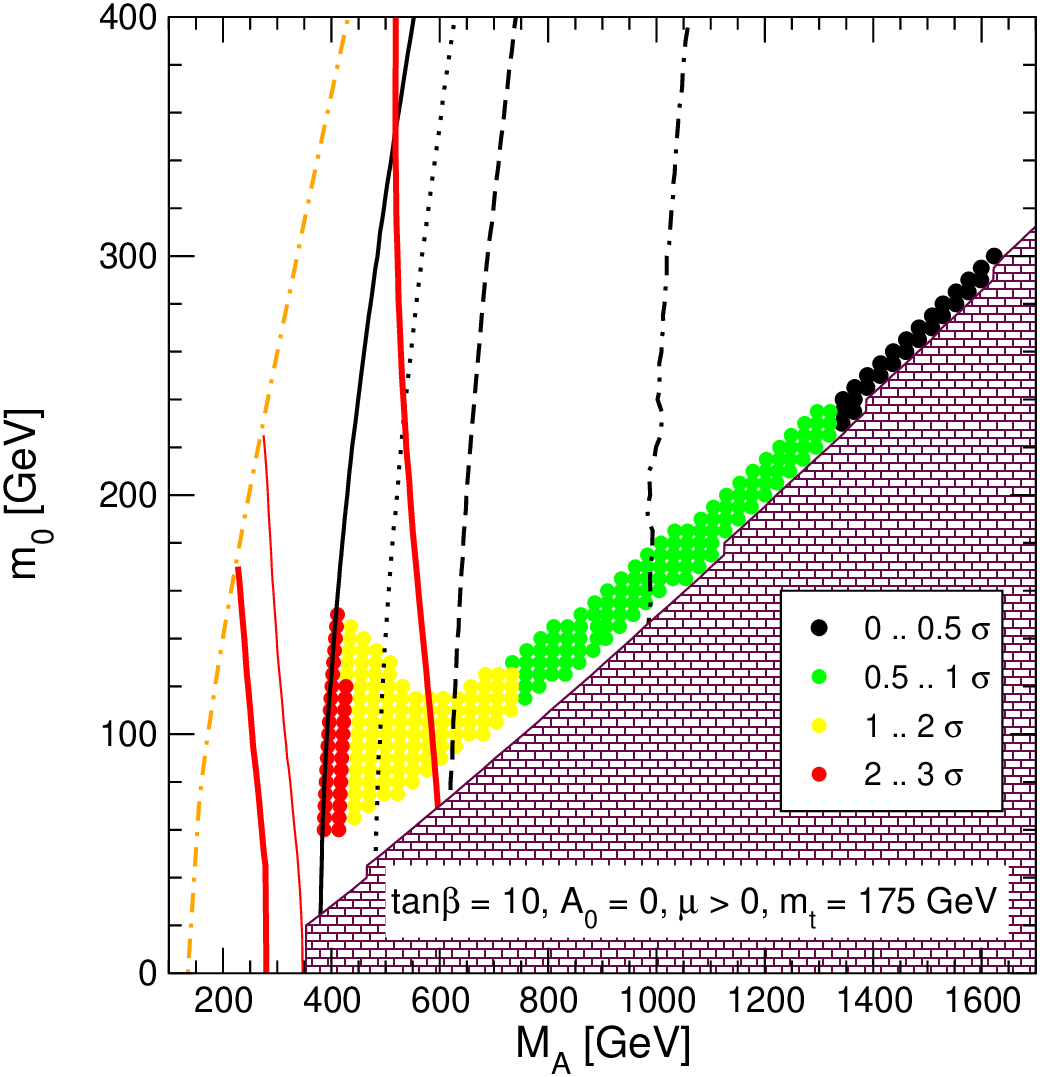,height=3.2in}
\epsfig{file=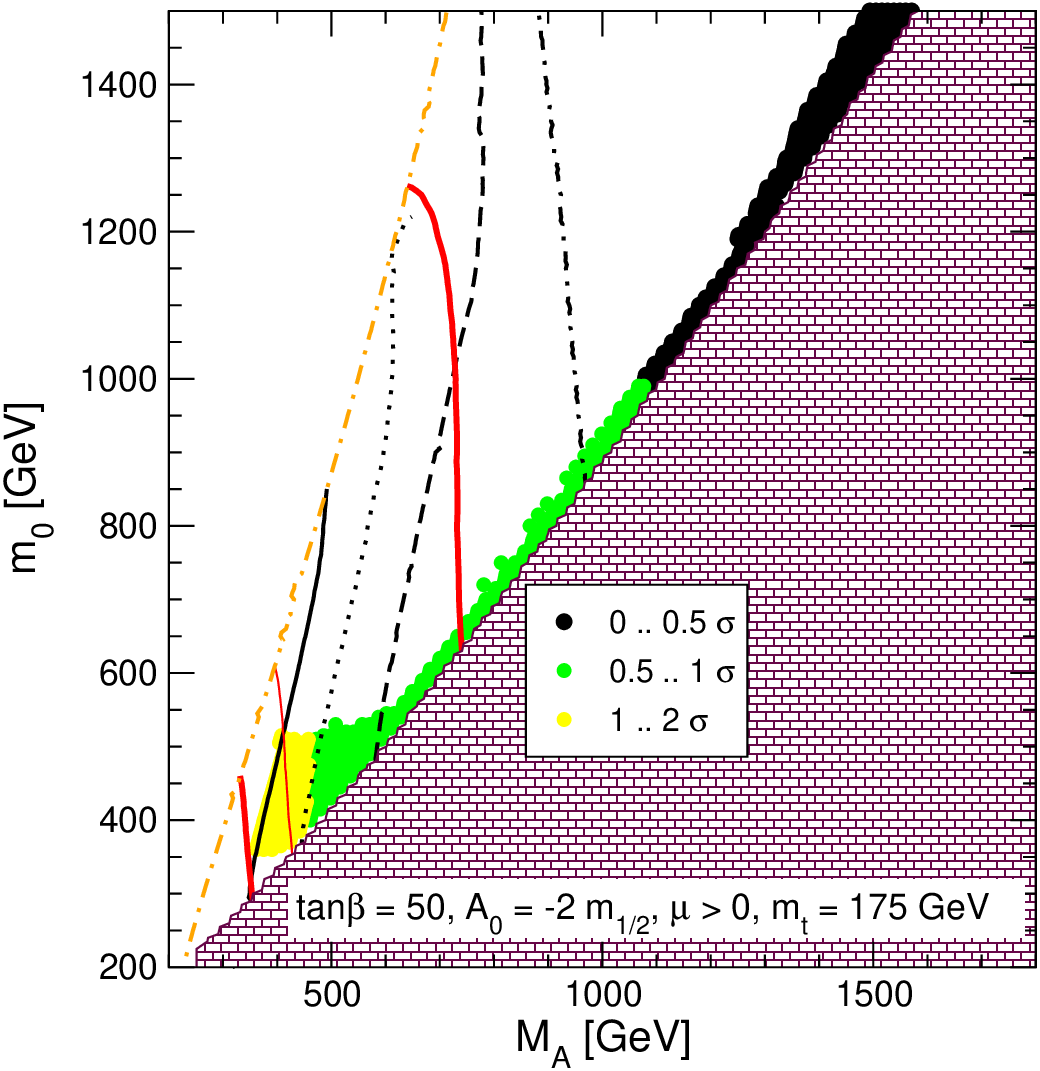,height=3.2in} \hfill
\end{minipage}

\vspace{.2em}
\begin{center}
\sibrfrac{\eeZh}{\hWW}
\end{center}

\begin{minipage}{8in}
\hspace{4cm} (c) \hspace{7.2cm} (d) 
\end{minipage}

\vspace{.2em}

\begin{minipage}{8in}
\epsfig{file=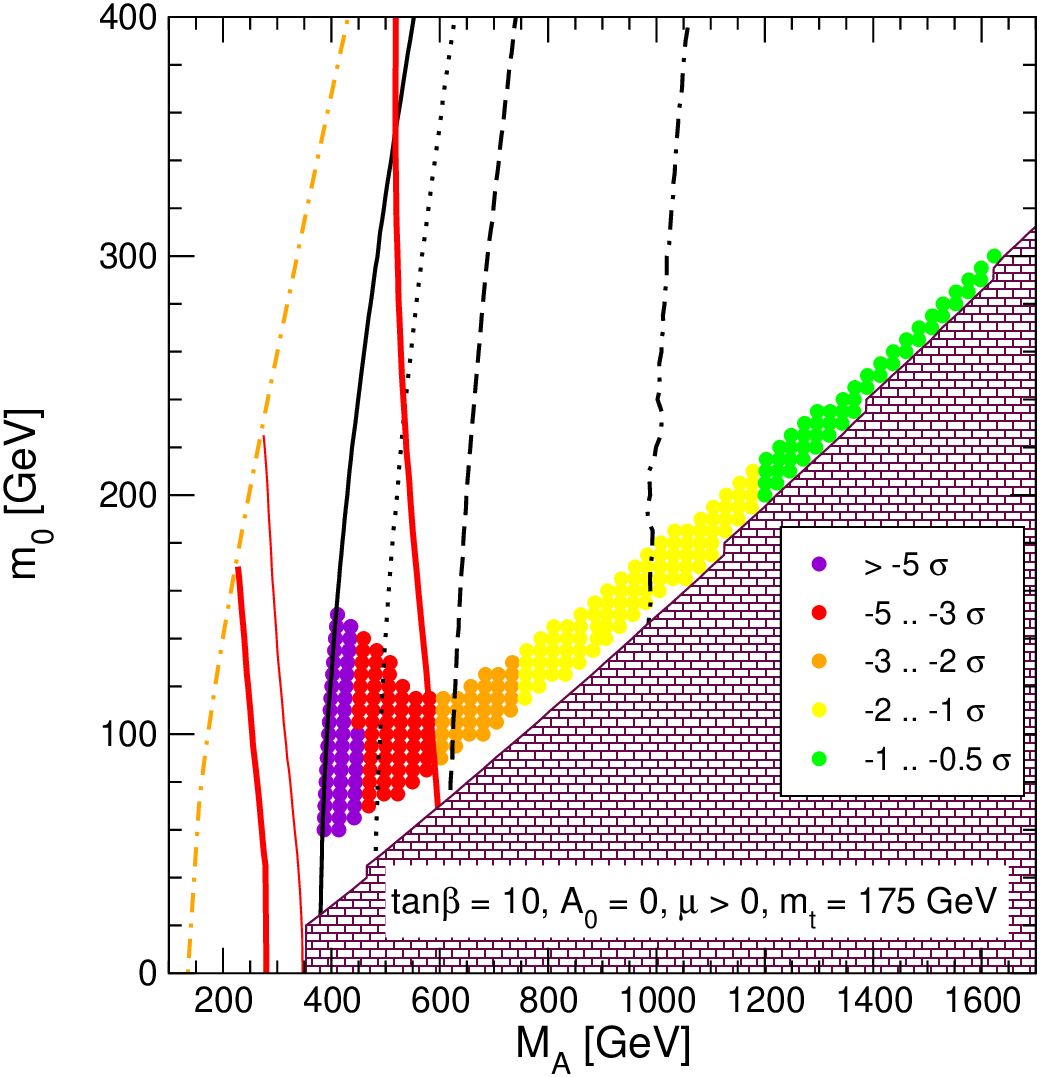,height=3.2in}
\epsfig{file=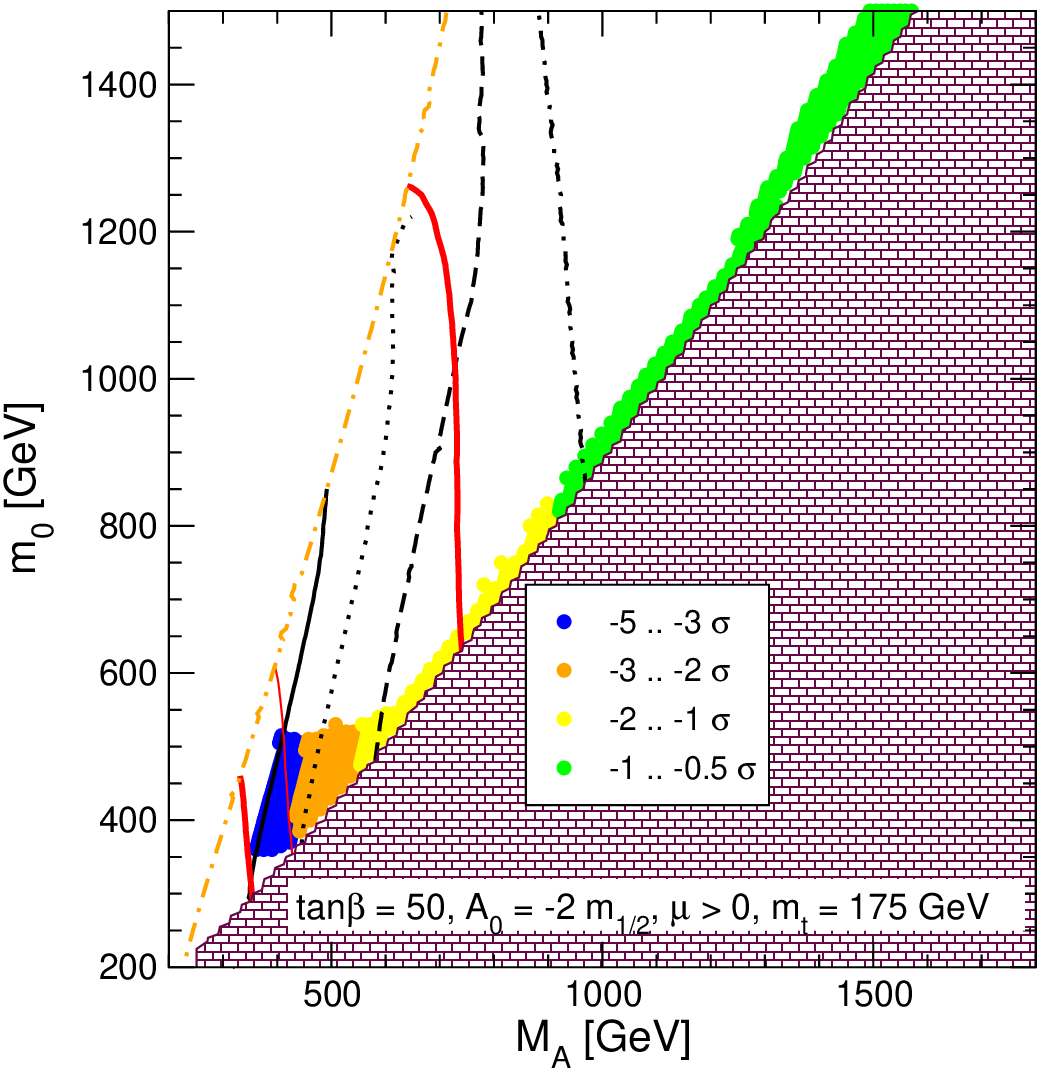,height=3.2in} \hfill
\end{minipage}

\caption{\it\small 
The deviations of \sibr{\eeZh}{\hbb} (upper row) and
\sibr{\eeZh}{\hWW} (lower row) for the lightest 
$\cp$-even CMSSM Higgs boson, normalized to the value in the SM with the
same Higgs mass, are given in the
$(\MA, m_0)$ planes for $\mu > 0$ and the values of $\tb$ and $A_0$
specified in the plots. The dot-dashed (orange) line represents the
border of the CMSSM parameter space.
The other contours and shadings are the same as in Fig.~\ref{fig:LCbb}.
\label{fig:LCMAreach}
}
\end{figure}

Our results are shown in more detail in \reffi{fig:LCMAreach}. The
deviations of the CMSSM predictions for \sibr{\eeZh}{\hbb} (upper row)
and \sibr{\eeZh}{\hWW} (lower row)
from the SM values are shown now in the ($\MA, m_0$) plane. 
For given values of $A_0$, $\tb$ and the sign of $\mu$, the allowed
interval for $\MA$ in the CMSSM compatible with a certain deviation of 
$\si \times \cB$ from the SM value can be read off directly. If one also
has information on $m_0$ and $m_{1/2}$, this can be compared with the 
value predicted for $\MA$ within the CMSSM. 

Within the NUHM, the indirect constraints on
$\MA$ obtained from the Higgs sector observables are analogous to the
constraints on the SM Higgs from the electroweak precision data. 
A direct measurement of the value of $\MA$ itself will provide a 
thorough consistency check of both the CMSSM and the NUHM.

The situation at the \gaC\ is somewhat worse than at the LC. Only for
small $\tb = 10$ can indirect limits on $\MA$ be obtained. 
In the $\hbb$ channel, limits of $\MA \lsim 900 \; (500) \gev$ can be
derived at the 1--2$\si$ level. Using the $\hWW$ channel,
corresponding limits of $700 \; (500) \gev$ might be possible.
Improved sensitivity could be obtained by combining LC and \gaC\ results, 
but such a combination goes beyond the scope of this paper.



\begin{figure}[!ht]
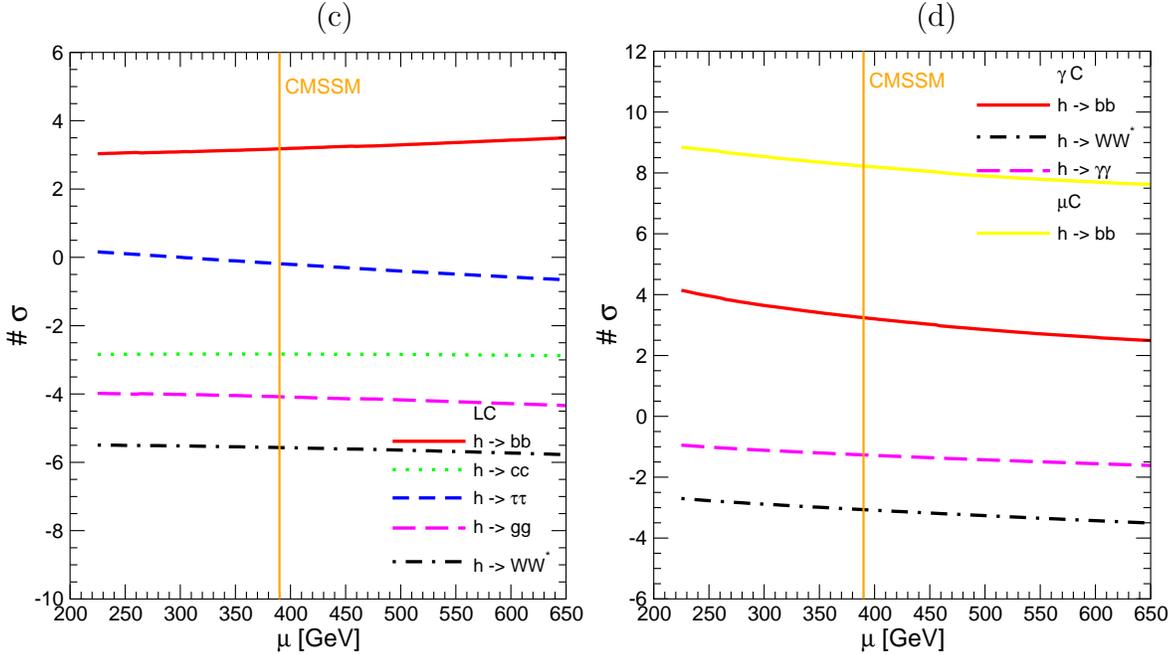


\begin{center}
Variation of the $\si \times \cB$ with $\MA$
\end{center}

\vspace{-1em}

\begin{minipage}{8in}
\hspace{4cm} (a) \hspace{7.2cm} (b) 
\end{minipage}

\vspace{.2em}

\begin{minipage}{8in}
\epsfig{file=EHOW2.NUHM_MA01.cl.eps,height=3.2in}
\epsfig{file=EHOW2.NUHM_MA02.cl.eps,height=3.2in} \hfill
\end{minipage}

\vspace{2em}
\begin{center}
Variation of the $\si \times \cB$ with $\mu$
\end{center}

\begin{minipage}{8in}
\hspace{4cm} (c) \hspace{7.2cm} (d) 
\end{minipage}

\vspace{.2em}

\begin{minipage}{8in}
\epsfig{file=EHOW2.NUHM_mu01.cl.eps,height=3.2in}
\epsfig{file=EHOW2.NUHM_mu02.cl.eps,height=3.2in} \hfill
\end{minipage}

\caption{\it\small 
The numbers of standard deviations of the predictions in the NUHM as
compared to the SM are shown in the different $\si \times \cB$ channels
for the LC (left column) and the \gaC\ and \muC\ (right column)
as functions of $\MA$ (upper row) and $\mu$ (lower row). 
The corresponding
CMSSM values of $\MA$ and $\mu$ are indicated by light vertical (orange) 
lines. The other
parameters have been chosen as $m_{1/2} = 300 \gev$, $m_0 = 100 \gev$, 
$\tb = 10$ and $A_0 = 0$.
\label{fig:NUHM}
}
\vspace{-1em}
\end{figure}

The analysis above demonstrates that 
the sensitivity of the Higgs sector observables to variations in $\MA$
can be useful for testing the universality assumption of the CMSSM,
i.e.\ for distinguishing between the CMSSM and the NUHM. In order to
investigate this issue in more detail, we 
focus on the NUHM scenario presented in Fig.~1 of~\cite{CMSSMuni}, in 
which $m_{1/2} = 300 \gev$, $m_0 = 100 \gev$,    
$\tb = 10$ and $A_0 = 0$ were chosen.  For this choice of parameters,
consistent models 
require $\MA \gsim 200 \gev$ and $|\mu| \lsim 650 \gev$, 

We show in \reffi{fig:NUHM} the values of $\si \times \cB$ (as before,
normalized to the SM value) obtained by
varying either $\MA$ or $\mu$ around the CMSSM point. The upper row shows
the variations with $\MA$, which may be substantial, particularly in the
$\hbb$ and $\hWW$ channels at the LC. The variations at the \gaC\ are
somewhat smaller, and those at the \muC\ could be larger (though this
information will presumably be available only on a longer time scale). 
In the case of the LC, the deviation from the CMSSM prediction 
(for which $\MA = 440 \gev$) could be as large as $\sim - 2.5 (+
1.5)\si$ or more in the $\hWW$ channel for $\de\MA = -(+) 100 \gev$.
The $\hbb$ channel is somewhat less
sensitive, with deviations of $\sim +1.5 (- 0.8)\si$ for the same range of
$\de\MA$. Thus, this key parameter of the MSSM Higgs boson sector, which
we recall is difficult to observe in this mass range, might be determined
indirectly within the framework of the NUHM. This would provide a
possibily for distinguishing between the CMSSM and the NUHM, even in the
case where no direct information on $\MA$ is available. 

The variation with $\mu$ (which in contrast to $\MA$ enters the Higgs 
sector observables only at loop level), as shown in the lower part of 
\reffi{fig:NUHM},
is much smaller than the $\MA$ variation. The deviations from the CMSSM
point of $\mu = 390 \gev$ barely exceed $1\si$ for all channels over the
entire parameter space. Therefore, in a scenario with relatively small
$\tb$, no substantial limits on $\mu$ can be
inferred from a precise measurement of the $\si \times \cB$ observables,
and one has to rely on direct measurements. In the case of $\mu$, this
will most likely be possible due to measurements in the gaugino sector of
the MSSM~\cite{gaugino}.


\section{Conclusions}
\label{sec:conclusions}

Extending our previous work on hadron colliders, we have analyzed in this
paper the abilities of $e^+ e^-$, $\ga \ga$ and $\mu^+ \mu^-$ colliders to 
constrain
MSSM parameters indirectly via accurate measurements of the production and
decays of the lightest $\cp$-even MSSM Higgs boson. We have estimated the
numbers of standard deviations by which $e^+ e^-$, $\ga \ga$ and $\mu^+
\mu^-$ measurements might differ from their
SM values, and shown how these
sensitivities vary with the CMSSM parameters $m_{1/2}, m_0, \tb$ and $A_0$.
We have shown that this information is potentially complementary to that
provided by direct searches for MSSM particles, as well as the indirect
constraints provided by $b \to s \ga$ decay, $g_{\mu} - 2$,
and cosmology. 

In particular,
we have shown that $e^+ e^-$, $\ga \ga$ and $\mu^+ \mu^-$ measurements
might be able to constrain the mass of the $\cp$-odd MSSM Higgs boson if
it remains undetected at the LHC or the LC. 
We have discussed the impact of the indirect information on $\MA$ both
within the framework of the CMSSM (where the mass of the $\cp$-odd MSSM 
Higgs boson is related to the CMSSM parameters $m_{1/2}, m_0, \tb$ and
$A_0$), and in the NUHM (in which the soft supersymmetry-breaking masses 
of the MSSM Higgs multiplets are allowed to differ from those of the 
squarks and sleptons). A direct observation of the $\cp$-odd MSSM Higgs 
boson, on the other hand, would enable stringent consistency tests both 
of the CMSSM and the NUHM.

We have furthermore demonstrated in this context that $e^+ e^-$, $\ga \ga$ 
and $\mu^+ \mu^-$ measurements are sensitive to deviations between the
CMSSM and the NUHM, i.e.\ they allow one to test the universality 
assumption
of the CMSSM. This refers in particular to deviations arising from
changing $\MA$ compared to its CMSSM value, while the sensitivities to
deviations from the CMSSM value of $\mu$, which enters the Higgs sector
observables only via loop corrections, are less promising.

We have emphasized in this paper the role of the Higgs-sector
observables and the indirect constraints from $b \to s \ga$, $g_{\mu} -2$,
and cosmology for testing supersymmetric models. In a realistic scenario
one would of course seek to combine this information with that
obtained from the possible observation of a spectrum of supersymmetric
particles, taking into account all available results from different
colliders. On the basis of the combined information obtained in this way 
one would then try to disentangle the detailed structure of
supersymmetry breaking.

The analysis performed in this paper, in which we have investigated the
sensitivity to deviations between two particular models, the CMSSM and
the NUHM, is a step into this direction, but there is clearly more work
needed along those lines. This could for instance involve a more
detailed exploration of the NUHM as well as models beyond the NUHM,
e.g., by relaxing further the universality assumptions for
the soft supersymmetry-breaking masses of squarks and sleptons.
We believe that this paper lays the basis for such further studies, by
quantifying the abilities of different colliders to constrain indirectly
MSSM parameters that may be difficult to measure directly at the LHC, or
even at a LC. Further work on the capabilities of $e^+ e^-$, $\ga \ga$ and
$\mu^+ \mu^-$ measurements to test supersymmetric models is clearly in
order to be prepared for the many different possibilities in which
supersymmetry might manifest itself in nature.


\subsection*{Acknowledgments}

\noindent  
The work of K.A.O.\ was partially
supported by DOE grant DE--FG02--94ER--40823.
This work has been supported by the European Community's Human
Potential Programme under contract HPRN-CT-2000-00149 Physics at
Colliders.




\begin{thebibliography}{99}

\bibitem{ehow} J.~Ellis, S.~Heinemeyer, K.~Olive and G.~Weiglein,
               {\em Phys. Lett.} {\bf B 515} (2001) 348, 
               hep-ph/0105067.

\bibitem{asbs} S.~Ambrosanio, A.~Dedes, S.~Heinemeyer, S.~Su and
               G.~Weiglein,
               {\em Nucl. Phys.} {\bf B 624} (2001) 3, 
               hep-ph/0106255.

\bibitem{robi} R.~Harlander and W.~Kilgore,
               {\em Phys. Rev. Lett.} {\bf 88} (2002) 201801,
               hep-ph/0201206;\\
               C.~Anastasiou and K.~Melnikov, hep-ph/0207004.

\bibitem{Battaglia:2001zp} M.~Battaglia et al.,
                           {\em Eur. Phys. J.} {\bf C 22} (2001) 535, 
                           hep-ph/0106204.


\bibitem{eennH} T.~Hahn, S.~Heinemeyer and G.~Weiglein,
                hep-ph/0211204.

\bibitem{teslatdr} TESLA TDR Part~3: {\it Physics at an $e^+e^-$ 
                   Linear Collider}, 
                   eds. R.D.~Heuer, D.~Miller, F.~Richard and P.M.~Zerwas, 
                   hep-ph/0106315,
                   see: {\tt http://tesla.desy.de/tdr} .

\bibitem{talkbrient} J.~Brient, 
                talk at the Linear Collider Workshop, Cracow, Poland,
                September 2001,\\
                see: {\tt http://webnt.physics.ox.ac.uk/lc/ecfadesy}.

\bibitem{gaC1} TESLA TDR Part~6, Chapter~1: 
               {\it Photon collider at TESLA},
               hep-ex/0108012, 
               see: {\tt http://tesla.desy.de/tdr} .

\bibitem{gaC2} D.~Asner et al.,
               hep-ex/0111056.

\bibitem{muC1} C.~Bl\"ochinger et al.,
               Higgs factory working group of the ECFA-CERN study on 
               Neutrino Factory \& Muon Storage Rings at CERN, 
               {\it Physics opportunities at $\mu^+\mu^-$ Higgs factories},
               hep-ph/0202199.

\bibitem{hff} S.~Heinemeyer, W.~Hollik and G.~Weiglein, 
              {\em Eur. Phys. Jour.} {\bf C 16} (2000) 139, 
              hep-ph/0003022.

\bibitem{hdecay} A.~Djouadi, J.~Kalinowski and M.~Spira,
                 {\em Comput. Phys. Commun.} {\bf 108} (1998) 56, 
                 hep-ph/9704448.

\bibitem{SUSYsearches} Particle Data Group, K. Hagiwara et al., 
                       {\em Phys. Rev.} {\bf D 66} (2002) 010001.

\bibitem{LEPHiggs} LEP Higgs working group, LHWG Note/2002-01,\\
                   {\tt http://lephiggs.web.cern.ch/LEPHIGGS/papers/}.

\bibitem{bsgexp} M.~Alam et al. [CLEO Collaboration],
                 {\em Phys. Rev. Lett.} {\bf 74} (1995) 2885, as updated in
                 S.~Ahmed et al., {CLEO CONF 99-10};\\
  $[$Belle Collaboration$]$, BELLE-CONF-0003, contribution to the 30th 
  International conference on High-Energy Physics, Osaka, 2000.
  See also\\
                 K.~Abe et al. [Belle Collaboration],
                 hep-ex/0103042;\\
                 K.~Abe {\it et al.},  [Belle Collaboration],
                 hep-ex/0107065; \\
                 L.~Lista  [BaBar Collaboration],
                 hep-ex/0110010.


\bibitem{bsgtheory} G. Degrassi, P. Gambino and G.~F. Giudice,
                    {\em JHEP} {\bf 0012} (2000) 009,
                    hep-ph/0009337; 
                    M.~Carena, D.~Garcia, U.~Nierste and C.~E.~Wagner,
                    {\em Phys. Lett.} {\bf B 499} (2001) 141,
                    hep-ph/0010003; \\
                    D.~Demir and K.~Olive,
                    {\em Phys. Rev.} {\bf D 65} (2002) 034007, 
                    hep-ph/0107329.


\bibitem{g-2} H.~Brown et al. [Muon $g_\mu - 2$ Collaboration],
              {\em Phys. Rev. Lett.} {\bf 86} (2001) 2227,
              hep-ex/0102017;  \\
              G.~Bennett et al. [Muon $g_\mu - 2$ Collaboration],
              {\em Phys. Rev. Lett.} {\bf 89} (2002) 101804
              [Erratum-ibid.\  {\bf 89} (2002) 129903],
              hep-ex/0208001.


\bibitem{EHNOS} J.~Ellis, J.~Hagelin, D.~Nanopoulos, K.~Olive
                and M.~Srednicki, 
                {\em Nucl. Phys.} {\bf B 238} (1984) 453; see also\\
                H.~Goldberg, 
                {\em Phys. Rev. Lett.} {\bf 50} (1983) 1419.

\bibitem{cdmexp} A.~Melchiorri and J.~Silk,
                 {\em Phys. Rev.} {\bf D 66} (2002) 041301, 
                 astro-ph/0203200.

\bibitem{CMSSMuni} J.~Ellis, K.~Olive and Y.~Santoso,
                   {\em Phys. Lett.} {\bf B 539} (2002) 107, 
                   hep-ph/0204192.

\bibitem{fh} S.~Heinemeyer, W.~Hollik and G.~Weiglein,
             {\em Comput. Phys. Commun.} {\bf 124} (2000) 76,
             hep-ph/9812320; 
             hep-ph/0002213;
             the {\tt FeynHiggs} code is available from \\
             {\tt http://www.feynhiggs.de}.

\bibitem{mhiggsFD2l} S.~Heinemeyer, W.~Hollik and G.~Weiglein,
                     {\em Phys. Rev.} {\bf D 58} (1998) 091701,
                     hep-ph/9803277;
                     {\em Phys. Lett.} {\bf B 440} (1998) 296,
                     hep-ph/9807423;
                     {\em Eur. Phys. Jour.} {\bf C 9} (1999) 343,
                     hep-ph/9812472.

\bibitem{mhiggsAEC} G.~Degrassi, S.~Heinemeyer, W.~Hollik, P.~Slavich
                    and G.~Weiglein,
                    {\em in preparation}. 

\bibitem{feynhiggs1.2} M.~Frank, S.~Heinemeyer, W.~Hollik and G.~Weiglein,
                       hep-ph/0202166.

\bibitem{mhiggsalt2} A.~Brignole, G.~Degrassi, P.~Slavich and F.~Zwirner,
                     {\em Nucl. Phys.} {\bf B 631} (2002) 195,
                     hep-ph/0112177.

\bibitem{mhiggsalsalb} A.~Brignole, G.~Degrassi, P.~Slavich and F.~Zwirner,
                       {\em Nucl. Phys.} {\bf B 643} (2002) 79, 
                       hep-ph/0206101.

\bibitem{EFGOSi} J.~Ellis, T.~Falk, G.~Ganis, K.~Olive and M.~Srednicki,
                 {\em Phys. Lett.} {\bf B 510} (2001) 236,
                 hep-ph/0102098.

\bibitem{g-2lbl} M.~Knecht and A.~Nyffeler,
                 {\em Phys. Rev.} {\bf D 65} (2002) 073034, 
                 hep-ph/0111058;\\
                 M.~Knecht, A.~Nyffeler, M.~Perrottet and E.~De~Rafael,
                 {\em Phys. Rev. Lett.} {\bf 88} (2002) 071802, 
                 hep-ph/0111059;\\
                 I.~Blokland, A.~Czarnecki and K.~Melnikov,
                 {\em Phys. Rev. Lett.} {\bf 88} (2002) 071803, 
                 hep-ph/0112117;\\
                 M.~Ramsey-Musolf and M.~Wise,
                 {\em Phys. Rev. Lett.} {\bf 89} (2002) 041601, 
                 hep-ph/0201297. 

\bibitem{Davier} M.~Davier, S.~Eidelman, A.~Hocker and Z.~Zhang,
                 hep-ph/0208177.

\bibitem{others} K.~Hagiwara, A.~Martin, D.~Nomura and T.~Teubner,
                 hep-ph/0209187; \\
                 F.~Jegerlehner, unpublished, as reported in
                 M.~Krawczyk,
                 hep-ph/0208076.


\bibitem{Melchiorri} A.~Benoit et al. [Archeops Collaboration],
                     astro-ph/0210306.

\bibitem{atlastdr} ATLAS Collaboration, 
  {\em Detector and Physics Performance Technical Design Report},
  CERN/LHCC/99-15 (1999), see:\\
 {\tt atlasinfo.cern.ch/Atlas/GROUPS/PHYSICS/TDR/access.html} .

\bibitem{LHHiggsProcs2001} D.~Cavalli et al.,
                           [Les Houches Higgs working group],
                           {\it Summary report}, 
                           hep-ph/0203056.

\bibitem{MAindirect} K.~Babu and C.~Kolda,
                     {\em Phys. Lett.} {\bf B 451} (1999) 77,
                     hep-ph/9811308;\\
                     M.~Battaglia and K.~Desch, hep-ph/0101165;\\
                     S.~Heinemeyer and G.~Weiglein, 
                     hep-ph/0102117;\\
                     J.~Guasch, W.~Hollik and S.~Pe\~naranda,
                     {\em Phys. Lett.} {\bf B 515} (2001) 367,
                     hep-ph/0106027;\\
                     M.~Carena, H.~Haber, H.~Logan and S.~Mrenna,
                     {\em Phys. Rev.} {\bf D 65} (2002) 055005, 
                     hep-ph/0106116;\\
                     A.~Curiel, M.~Herrero, D.~Temes and J.~De Troconiz,
                     {\em Phys. Rev.} {\bf D 65} (2002) 075006,
                     hep-ph/0106267;\\
                     S.~Dawson and S.~Heinemeyer,
                     {\em Phys. Rev.} {\bf D 66} (2002) 055002,
                     hep-ph/0203067.

\bibitem{gigaz} J.~Erler, S.~Heinemeyer, W.~Hollik, G.~Weiglein 
                and P.M.~Zerwas,
                {\em Phys. Lett.} {\bf B 486} (2000) 125,
                hep-ph/0005024.

\bibitem{EFOS} J.~Ellis, T.~Falk, K.~Olive and Y.~Santoso,
               hep-ph/0210205.

\bibitem{gaugino} S.~Choi, J.~Kalinowski, G.~Moortgat-Pick and P.~Zerwas,
                  {\em Eur. Phys. J.} {\bf C 22} (2001) 563, 
                  hep-ph/0108117;
                  hep-ph/0202039.


\end{thebibliography}
\end{document}